# Pilot Contamination for Wideband Massive MMO: Number of cells Vs Multipath (Draft With More Detailed Derivation and Results)

Tadilo Endeshaw Bogale, *Member, IEEE*, Long Bao Le, *Senior Member, IEEE*, Xianbin Wang *Senior Member, IEEE* and Luc Vandendorpe *Fellow, IEEE*

*Abstract*—This paper proposes a novel joint channel estimation and beamforming approach for multicell wideband massive multiple input multiple output (MIMO) systems. With the proposed channel estimation and beamforming approach, we determine the number of cells $N_c$ that can utilize the same time and frequency resource while efficiently mitigating the effect of pilot contamination. The proposed approach exploits the multipath characteristics of wideband channels. Specifically, when the channel has $L$ multipath taps, it is shown that $N_c \leq L$ cells can reliably estimate the channels of their user equipments (UEs) and perform beamforming while mitigating the effect of pilot contamination. In an exemplify setting, for a long term evolution (LTE) channel environment having delay spread $T_d = 4.69\mu$ second and channel bandwidth $B = 2.5$**MHz**, we have found that $L = 18$ cells can use this band. In practice, $T_d$ is constant for a particular environment and carrier frequency, and hence $L$ increases as the bandwidth increases. All the analytical expressions have been validated, and the superiority of the proposed design is demonstrated using extensive numerical simulations both for correlated and uncorrelated channels. The proposed channel estimation and beamforming design is linear and simple to implement.

## I. INTRODUCTION

Massive multiple input multiple output (MIMO) technology is one of the promising means for achieving very high energy and spectrum efficiency requirements of the future 5G networks [1]–[3]. It is anticipated that the 5G network could employ massive MIMO systems both at microwave and millimeter wave (mmWave) frequency bands where beamforming is their key ingredient. The beamforming gain envisaged by the massive MIMO system depends on the availability of full and accurate channel state information (CSI). However, in practice, the channel between the transmitter and receiver is estimated from orthogonal pilot sequences which are limited by the coherence time of the channel [4], [4]–[7]. In a multiuser and multicell setup, massive MIMO can be deployed both at the base stations (BSs) and mobile stations (UEs).

However, deploying massive number of antennas at the UEs is usually infeasible from practical viewpoint particularly at microwave frequency bands. This is because the deployment of massive antenna requires sufficient spacing between antennas in order to get the expected multiplexing and diversity gains of a MIMO channel. Furthermore, the energy consumption and cost of the transceiver device increase as the number of antennas increases. For these reasons, it is economical to deploy antenna arrays at the BSs and single antennas at each UE for microwave frequency band applications. This motivates us to consider a massive MIMO system where each UE has single antenna as in [4].

Channel estimation can be performed either in frequency division duplex (FDD) or time division duplex (TDD) approaches. In an FDD approach, first, the transmitters send pilot sequences to all receivers. Then, each receiver will estimate its own channel. Finally, the estimated channel is fed back to the transmitters via feedback channel. In a TDD based approach, however, the channel coefficients can be estimated at the transmitters, receivers or both appropriately. In the conventional TDD and FDD based channel estimation approaches, the pilot sequences needs to be orthogonal to maintain sufficient channel estimation quality which are directly related to the number of antennas where the pilot signals are transmitted (i.e., the required pilot sequences increase as the number of antennas increase). Thus, for a massive MIMO system, one can possibly estimate and exploit the maximum number of channel coefficients using TDD approach by sending pilots from UEs. By doing so, each BS will estimate and utilize the channel coefficients both for the downlink and uplink channel transmissions. In a multicell setup where no coordination takes place between cells, each BS independently determines its number of UEs from the coherence time of the channel only. Furthermore, as the number of orthogonal pilot sequences is limited for a given coherence time and bandwidth, the UEs in each cell are forced to utilizes the same (correlated) pilot sequences. The use of correlated pilot sequences for several co-channel cells will create a so called *pilot contamination* [4][1]. It is shown that pilot contamination will limit the achievable signal to interference plus noise ratio (SINR) of each UE to a finite value irrespective of the transmission power and number of antennas. For this reason, pilot contamination creates a performance bottleneck in terms of channel capacity for multicell massive MIMO systems [1], [4]–[10].

---
[1]We would like to emphasize that it is always possible to increase the number of UEs per cell just by orthogonalizing UEs over different subcarriers. However, such an approach will not fundamentally improve the overall network throughput (see Section VII of [4] for details). For this reason, we assume that each UE employs all the available bandwidth as in [4].

A number of approaches have been proposed to address pilot contamination which can be broadly classified as: (i). Sub-space method which exploits the sub-space information of the channel covariance matrices [5], [7], [11] (ii). Pilot optimization/scheduling/shifting method which optimizes/schedules/shifts the pilot sequences of each cell (user) while utilizing appropriate signal dimensions such as time and frequency [6], [9], [12], [13] (iii). Pilot contamination beamforming (precoding) approach that performs beamforming while taking into account the effect of pilot contamination [14]. (iv). A method exploits the transmitted signal modulation information [15] which are briefly summarized as follows. Eigenvalue decomposition (EVD) based channel estimation approach is proposed in [7]. In this paper, it is demonstrated that the EVD based channel estimation method achieves better estimation accuracy than that of the conventional linear channel estimation method. A coordinated channel estimation approach is proposed in [5] to eliminate (possibly mitigate) pilot contamination. This paper shows that the effect of pilot contamination can be eliminated by selecting/scheduling UEs having particular channel covariance matrices. In [11], a subspace based channel estimation is proposed to improve the channel estimation of massive MIMO systems.

In [8], pilot contamination elimination approach is proposed by allowing pilot transmissions both in the uplink and downlink channels. In [9], successive pilot transmission approach has been proposed to address pilot contamination issues. The approach of this paper utilizes consecutive pilot transmission phases in which each BS stays idle at one phase and repeatedly transmits pilot sequences in other phases. In [12], [13], time shifted channel estimation approach for orthogonal frequency division multiple access (OFDMA) based frequency selective channels is proposed. In such an approach, first the cells are grouped in some predefined way. Then, from these groups, only one group of cells sends pilot signal sequentially while all the other groups of cells transmit their downlink data signals. Finally, all cells transmit their uplink data on the remaining time slot. During the pilot transmission phase of a group of cells, since the data signals in all the other groups take place in the downlink channel, the interference signal arises between BSs which can ultimately be mitigated when the number of antennas at each BS tends to infinity. Due to this reason, pilot contamination will not occur between groups of cells. However, since the number of cells in each group could be more than one, larger coherence time is required to efficiently mitigate the effect of pilot contamination in each group of cells, and to allow large number of groups when each cell serves the same number of UEs like in [4]. In [6], it is shown that the channel estimation accuracy can be improved by optimizing the pilot sequences based on their channel covariance information when the channel coherence time is smaller than the total number of UEs.

In [14], linear minimum mean square error (LMMSE)-based precoding method that mitigates pilot contamination problem is provided. The approach of this method designs the precoding matrices of each cell such that the sum of the squared error of its own UEs and interference created to UEs of all other cells is minimized (see also [10]). In [15], the problem of pilot contamination has been addressed for wideband channels by considering that the transmitter applies cosine modulated multitone (CMT) transmission scheme. By doing so, the equalization process at the receiver can be performed blindly which consequently removes the channel estimation errors (due to pilot contamination effect) without any need for cooperation between different cells. From the aforementioned discussions, one can understand that the problem of pilot contamination has been examined by several existing works. However, some of the aforementioned papers address pilot contamination by exploiting particular covariance information or modulation schemes which may not hold in all scenarios. The remaining approaches utilize extra resources compared to [4] or introduce additional assumptions that may not hold true for all practically relevant settings for reducing or eliminating pilot contamination (for example, the works of [5]–[7] are not able to mitigate the effect of pilot contamination when UEs have the same channel covariance matrix).

The current paper considers the pilot contamination problem for multicell massive MIMO systems with frequency selective channels. For a given pilot duration, we assume that the number of UEs served by each BS is the same as that in [4], [12]. Furthermore, the channel covariance matrix of each UE is assumed to have "arbitrary structure". Under these settings and assumptions, we propose a novel joint channel estimation and beamforming approach. The proposed approach exploits the inherent multipath behavior of frequency selective channels. Specifically, when the channel has a maximum of $L$ multipath taps, it is shown that $N_c \leq L$ cells can reliably estimate the channels of their UEs and perform beamforming while efficiently mitigating the effect of pilot contamination. In other words, when $N_c \leq L$, it is shown that the SINR of each UE's sub-carrier increases as the number of antennas at each BS $N$ increases (i.e., the SINR grows indefinitely as $N \to \infty$). As will be clear in Section V, this improvement is possible by reducing the net pilots used for each UE by a factor of $L$ compared to the existing designs (i.e., allocating only one pilot per UE for any $L$), and applying the proposed joint channel estimation and beamforming approach.

In a typical long term evolution (LTE) channel environment having delay spread $T_d = 4.69\mu s$ and channel bandwidth $B = 2.5$MHz, $L = 18$ cells can use this frequency band while efficiently mitigating the effect of pilot contamination. In practice, $T_d$ is constant for a particular environment and carrier frequency, and hence $L$ increases as the bandwidth increases. Furthermore, these $L$ multipath channel coefficients can be modeled as random variables [16] (see Appendix E for a concise description on the relation between $B$, $T_d$ and $L$). We would like to emphasize here that the current paper employs the statistical model of the $L$ multipath channel coefficients (see Section II). Hence, $L$ does not necessarily imply the non-zero elements of multipath channel taps in all channel realizations (i.e., a given multipath channel coefficient could be zero (closer to zero) in one channel realization and nonzero in another one). This validates that the assumption of the current paper is realistic. The proposed design is linear, simple to implement and significantly outperforms the existing designs, and is validated by extensive simulations.

TABLE I
FREQUENTLY USED VARIABLES, VECTORS AND MATRICES

| Var. | Definition |
|---|---|
| $\bar{\mathbf{h}}_{kjin}$ | Channel between $k$th UE in $j$th cell to $i$th BS $n$th antenna |
| $h_{kjins}$ | The $s$th sub-carrier channel corresponding to $\bar{\mathbf{h}}_{kjin}$ |
| $\hat{h}_{kiins}$ | The estimate of $h_{kiins}$ |
| $L$ | Number of multipath taps |
| $N_p$ | Number of pilot samples |
| $T_p$ | Pilot duration |
| $\tilde{K}$ | Number of UEs served at each cell |
| $T_d$ | Delay spread |
| $N_c$ | Number of cells |
| $\bar{\mathbf{H}}_{kji}$ | Channel matrix between $k$th UE in $j$th cell to $i$th BS |
| $\mathbf{P}_{kji}$ | Temporal correlation matrix of $\bar{\mathbf{H}}_{kji}$ |
| $\mathbf{R}_{kji}$ | Spatial correlation matrix of $\bar{\mathbf{H}}_{kji}$ |
| $\mathbf{x}_{kj}$ | Pilot symbols transmitted from the $k$th UE in cell $j$ |
| $\mathbf{C}_{mji}$ | $\mathrm{tr}\{\mathbf{R}_{mji}\}\mathbf{P}_{mji}^* = N\mathbf{P}_{mji}^*$ |
| $\mathbf{C}_{mjis}$ | $(\mathbf{f}_s^H \mathbf{P}_{mji}\mathbf{f}_s)\mathbf{R}_{mji}^*$ |
| $\mathbf{C}_{mjuvis}$ | $(\mathbf{f}_s^H \mathbf{P}_{uvi}\mathbf{f}_s)\mathrm{tr}\{\mathbf{R}_{uvi}^*\mathbf{R}_{mji}^*\}\mathbf{P}_{mji}^*$ |
| $\mathbf{v}_{kis}$ | Linear combination vector to compute $\hat{h}_{kiins}$ |
| $\mathbf{a}_{kiis}$ | Uplink beamforming vector of cell $i$ $k$th UE sub-carrier $s$ |
| $\mathbf{b}_{kiis}$ | Downlink precoding vector of cell $i$ $k$th UE sub-carrier $s$ |

This paper is organized as follows. Section II discusses the system and channel models. In Section III, a brief summary of pilot contamination and the objective of the paper is explained. In Section IV - VI, the proposed joint channel estimation and beamforming designs is provided. In Section VII, performance analysis and discussions on the proposed design is presented briefly. In Section VIII, extensive simulation results are presented. Finally, Section IX draws conclusions.

*Notations:* In this paper, upper/lower-case boldface letters denote matrices/column vectors. $\mathbf{X}_{(i,j)}$, $\mathbf{X}^T$, $\mathbf{X}^H$ and $\mathrm{E}(\mathbf{X})$ denote the $(i,j)$th element, transpose, conjugate transpose and expected value of $\mathbf{X}$, respectively. $\mathrm{diag}(.)$, $\mathrm{blkdiag}(.)$, $|.|$, $\lceil x \rceil$, $\mathbf{1}_N$, $\mathbf{I}$ and $\mathbb{C}^{N \times M}(\Re^{N \times M})$ denote diagonal, block diagonal, two norm, nearest integer greater than or equal to $x$, an $N$ sized vector of ones, appropriate size identity matrix and $N \times M$ complex (real) entries, respectively. The acronym s.t, i.i.d and ZMCSCG denote "subject to", "independent and identically distributed" and zero mean circularly symmetric complex Gaussian, respectively. For better readability of the paper, we have summarized the most frequently used scalars, vectors and matrices in Table I.

## II. SYSTEM AND CHANNEL MODEL

We consider a multiuser and multicell system with $N_c$ cells (i.e., BSs). It is assumed that each UE and BS are equipped with 1 and $N$ antennas, respectively. We consider a transmission scheme with symbol period $T_s$ and channel environment with maximum delay spread $T_d$[2]. The total transmission bandwidth becomes $B = B_0(1+\alpha)$, where $B_0 = \frac{1}{2T_s}$ is the Nyquist bandwidth and $\alpha$ is the excess bandwidth which has typical values $0.2 \leq \alpha \leq 0.35$ [19], [20].

---
[2]The transmission environment has a maximum delay spread of $T_d$ which has typical values $T_d = \{4.7, 5.2\}\mu s$ for urban cells and $T_d = 16.7\mu s$ for very large multi-cell systems in the existing LTE network [4], [17], [18].

For these settings, the number of multipath channel taps $L$ can be approximated as $L = \frac{T_d}{T_s}$ [20]. The multipath channel coefficients between the $k$th UE in $j$th cell to the $i$th BS $n$th antenna can be represented as [17], [20]

$$\bar{\mathbf{h}}_{kjin} = [\bar{h}_{kjin1}, \bar{h}_{kjin2}, \cdots, \bar{h}_{kjinL}]^T. \quad (1)$$

The analytical model for $\bar{\mathbf{H}}_{kji} = [\bar{\mathbf{h}}_{kji1}, \bar{\mathbf{h}}_{kji2}, \cdots, \bar{\mathbf{h}}_{kjiN}]$ may vary from one wireless standard to another. In the current paper, we model the $l$th channel tap (i.e., $\bar{h}_{kjinl}$) as a ZMCSCG random variable each with variance $\sigma_l^2$ as in [20] (see Chapter 2.4.2 of [20]) but these channel coefficients are correlated spatially over the BS antenna arrays. We further assume that the spatial correlation matrix of all multipath components are the same. Under such assumptions, we can express $\bar{\mathbf{H}}_{kji}$ as [21]

$$\bar{\mathbf{H}}_{kji} = \sqrt{\mathbf{P}_{kji}}\tilde{\mathbf{H}}_{kji}\sqrt{\mathbf{R}_{kji}} \quad (2)$$

where each entry of $\tilde{\mathbf{H}}_{kji}$ is characterized by an i.i.d ZMCSCG random variable with unit variance, $\mathbf{P}_{kji} = \mathrm{diag}(\sigma_{kji1}^2, \sigma_{kji2}^2, \cdots, \sigma_{kjiL}^2)$ is a diagonal matrix and $\mathbf{R}_{kji} \in \mathcal{C}^{N \times N}$ is the positive semidefinite spatial correlation matrix.

In most existing wideband systems having multipath channel coefficients, OFDM based transmission is adopted. For such a transmission, the channel coefficient of each sub-carrier has practical importance. In fact, these channel coefficients can be obtained from the linear combination of the multipath channel coefficients derived in (1). To this end, the channel between the $k$th UE in $j$th cell to the $i$th BS $n$th antenna in sub-carrier $s$ can be expressed as [17], [20]

$$h_{kjins} = \mathbf{f}_s^H \bar{\mathbf{h}}_{kjin} \quad (3)$$

where $\mathbf{f}_s^H = [1, e^{-\sqrt{-1}\frac{2\pi}{M}s}, e^{-\sqrt{-1}\frac{2\pi}{M}2s}, \cdots, e^{-\sqrt{-1}\frac{2\pi}{M}(L-1)s}]$ with $M$ as the fast Fourier transform (FFT) size of the OFDM transmission.

## III. PILOT CONTAMINATION AND OBJECTIVE

The number of UEs served by a BS is proportional to the pilot duration $T_p$. For fixed $T_p$, the channel coefficients of each UE can be estimated with or without OFDM pilot transmission. By denoting each OFDM duration as $T_o$ and useful symbol duration as $T_u$, each BS can serve the following number of UEs (see (4) of [4] and Section II of [12]):

$$\frac{T_p}{T_d}\frac{T_u}{T_o} \leq \frac{T_p}{T_d} = \frac{N_p}{L} \triangleq \tilde{K} \quad (4)$$

where $N_p = \frac{T_p}{T_s}$ is the number of samples acquired in $T_p$. If each UE applies non OFDM pilot transmission, it is possible to serve $\tilde{K}$ UEs [4].

If a BS serves $\tilde{K}$ UEs (i.e., use $T_p$ to estimate the channel coefficients of these UEs), we can understand from the work of [4] that only one cell can serve its UEs without experiencing any pilot contamination. In particular, when $\mathbf{P}_{kji} = g_{kji}\mathbf{I}$ and $\mathbf{R}_{kji} = \mathbf{I}$, and if there are multiple cells where the BS of each cell has sufficiently large number of antennas (i.e., $N \to \infty$),

the effect of noise vanishes and the SINR of the $k$th UE in cell $i$ and sub-carrier $s$ is given as

$$SINR_{kis} = \frac{\mathrm{E}\{h_{kiins}^H h_{kiins}\}}{\sum_{j \neq i}^{N_c} \mathrm{E}\{h_{kjins}^H h_{kjins}\}} = \frac{g_{kii}^2}{\sum_{j \neq i}^{N_c} g_{kij}^2}. \quad (5)$$

As we can see from this equation, $SINR_{kis}$ is bounded even if $N \to \infty$. However, one can understand from MIMO communication that increasing $N$ should help to improve $SINR_{kis}$ if pilot contamination is mitigated or cancelled. This motivates us to consider the following objective:

For the given $B$, $T_p$, $T_o$, $T_s$, $T_d$ and $T_u$, the channel vector between any two UEs are uncorrelated and each BS serves $\tilde{K}$ UEs (i.e., the same setting as in [4], [12])[3], how many cells can utilize the same time-frequency resource while ensuring that $SINR_{kis}$ increases as $N$ increases (i.e., achieving unbounded $SINR_{kis}$ as $N \to \infty$ which is the same as mitigating or possibly canceling the effect of pilot contamination)?

## IV. PROPOSED CHANNEL ESTIMATION AND BEAMFORMING

This section discusses the proposed channel estimation and beamforming approach. In this regard, we assume that the pilots are transmitted without OFDM whereas, the data symbols are transmitted using OFDM scheme. The current paper considers that both pilot and data transmissions take place in the uplink channel. We also assume that $N_c$ (which will be determined in the sequel) cells utilize the same time-frequency resource while mitigating the effect of pilot contamination. Here, we consider that the $i$th cell is the target cell.

### A. Channel Estimation

This subsection provides the proposed channel estimation. For the considered setting, the received signal at the $n$th antenna of the $i$th BS can be expressed as

$$\mathbf{r}_{in} = \sum_{k=1}^{\tilde{K}} (\mathbf{X}_{ki} \bar{\mathbf{h}}_{kiin} + \sum_{j=1, j \neq i}^{N_c} \mathbf{X}_{kj} \bar{\mathbf{h}}_{kjin}) + \mathbf{w}_{in}$$
$$= \mathbf{X}_i \bar{\mathbf{h}}_{iin} + \tilde{\mathbf{X}}_i \tilde{\mathbf{h}}_{iin} + \mathbf{w}_{in} \quad (6)$$

where $\bar{\mathbf{h}}_{jin} = [\bar{\mathbf{h}}_{1jin}^T, \bar{\mathbf{h}}_{2jin}^T, \cdots, \bar{\mathbf{h}}_{\tilde{K}jin}^T]^T$, $\tilde{\mathbf{h}}_{iin} = [\bar{\mathbf{h}}_{1in}^T, \bar{\mathbf{h}}_{2in}^T, \cdots, \bar{\mathbf{h}}_{(i-1)in}^T, \bar{\mathbf{h}}_{(i+1)in}^T, \cdots, \bar{\mathbf{h}}_{N_c in}^T]^T$, $\tilde{\mathbf{X}}_i = [\mathbf{X}_1, \mathbf{X}_2, \cdots, \mathbf{X}_{i-1}, \mathbf{X}_{i+1}, \cdots, \mathbf{X}_{N_c}] \in \mathcal{C}^{N_p \times (N_p(N_c-1))}$, $\mathbf{X}_i = [\mathbf{X}_{1i}, \mathbf{X}_{2i}, \cdots, \mathbf{X}_{\tilde{K}i}]$ and $\mathbf{w}_{in} \in \mathcal{C}^{N_p \times 1}$ is the additive noise vector during pilot transmission phase where its entries are assumed to be i.i.d ZMCSCG random variables each with variance $\sigma^2$, and $\mathbf{X}_{kj} \in \mathcal{C}^{N_p \times L}$ is a Toeplitz matrix, i.e.,

$$\mathbf{X}_{kj} = \quad (7)$$

$$\begin{bmatrix} x_{kj_1} & 0 & \cdots & 0 & 0 \\ x_{kj_2} & x_{kj_1} & \cdots & \vdots & \vdots \\ x_{kj_3} & x_{kj_2} & \cdots & 0 & 0 \\ \vdots & \vdots & \cdots & \vdots & \vdots \\ x_{kj_{(N_p-1)}} & x_{kj_{(N_p-2)}} & \cdots & x_{kj_{(N_p-L+1)}} & x_{kj_{(N_p-L)}} \\ x_{kj_{N_p}} & x_{kj_{N_p-1}} & \cdots & x_{kj_{(N_p-L+2)}} & x_{kj_{(N_p-L+1)}} \end{bmatrix}$$

with $\mathbf{x}_{kj} = [x_{kj_1}, x_{kj_2} \cdots, x_{kj_{N_p}}]^T$ as the pilot symbols transmitted from the $k$th UE in cell $j$ which will be designed in Section V. As can be seen from (6), the $i$th BS experiences pilot contamination due to the term $\tilde{\mathbf{X}}_i \tilde{\mathbf{h}}_{iin}$ [4]. For an arbitrary pilot, one can also notice from this equation that $\tilde{\mathbf{X}}_i \tilde{\mathbf{h}}_{iin} = 0$ is ensured when $N_c = 1$ (i.e., like in [4], [12]). For such setting, one can apply MMSE or LS method to estimate $\bar{\mathbf{h}}_{iin}$ from $\mathbf{r}_{in}$.

[3]We would like to mention here that we have taken the system setup in [4], [12] as it is commonly adopted, and for a fair comparison of the proposed and existing designs.

In the following, we provide the proposed channel estimation approach. In this regard, we introduce a vector $\mathbf{v}_{kis} \in \mathcal{C}^{N_p \times 1}$ to estimate the $s$th sub-carrier channel of the $k$th UE in cell $i$ which will be designed in Section V. Using this linear combination vector, we express the estimate of $h_{kiins}$ as

$$\hat{h}_{kiins} = \mathbf{r}_{in}^T \mathbf{v}_{kis} = (\mathbf{X}_i \bar{\mathbf{h}}_{iin} + \sum_{j=1, j \neq i}^{N_c} \mathbf{X}_j \bar{\mathbf{h}}_{jin} + \mathbf{w}_{in})^T \mathbf{v}_{kis}. \quad (8)$$

As will be clear in the next section, this estimated channel helps us to increase the number of cells more than one.

### B. Beamforming

As discussed above, we consider the uplink channel for data transmission using OFDM approach. Since data transmission takes place using OFDM technique, the received signal of each sub-carrier can be obtained independently at each antenna of all BSs. To this end, the $i$th BS $n$th antenna receives the following uplink signal in sub-carrier $s$ ($y_{ins}$)

$$y_{ins} = \sum_{k=1}^{\tilde{K}} \sum_{j=1}^{N_c} h_{kjins} d_{kjs} + \tilde{w}_{ins}$$

where $\tilde{w}_{ins}$ is the noise sample at the $i$th BS $n$th antenna and sub-carrier $s$ during data transmission and $d_{kjs}$ is the data symbol transmitted in sub-carrier $s$ of the $k$th UE in cell $j$. The overall received signal at the $i$th BS $s$th sub-carrier becomes

$$\mathbf{y}_{is} = \sum_{k=1}^{\tilde{K}} \sum_{j=1}^{N_c} \mathbf{h}_{kjis} d_{kjs} + \tilde{\mathbf{w}}_{is}$$
$$= \mathbf{h}_{kiis} d_{kis} + \sum_{m=1}^{\tilde{K}} \sum_{j=1, (m,j) \neq (k,i)}^{N_c} \mathbf{h}_{mjis} d_{mjs} + \tilde{\mathbf{w}}_{is}$$

where $\mathbf{y}_{is} = [y_{i1s}, y_{i2s}, \cdots, y_{iNs}]^T$ and $\tilde{\mathbf{w}}_{is} = [\tilde{w}_{i1s}, \tilde{w}_{i2s}, \cdots, \tilde{w}_{iNs}]^T$. It is assumed that each entry of $\tilde{\mathbf{w}}_{is}$ is assumed to be i.i.d ZMCSCG random variable with variance $\tilde{\sigma}^2$. Now, let us assume that we are interested in estimating $d_{kis}$ using a beamforming vector $\mathbf{a}_{kiis} \in \mathcal{C}^{N \times 1}$ as

$$\hat{d}_{kis} = \mathbf{a}_{kiis}^H \mathbf{y}_{is} \qquad (9)$$

$$= \mathbf{a}_{kiis}^H (\mathbf{h}_{kiis} d_{kis} + \sum_{m=1}^{\tilde{K}} \sum_{j=1,(m,j)\neq(k,i)}^{N_c} \mathbf{h}_{mjis} d_{mjs} + \tilde{\mathbf{w}}_{is}).$$

Using this beamformer, $\hat{d}_{kis}$ will have the following SINR

$$\bar{\gamma}_{kis} = \frac{\mathrm{E}|\mathbf{h}_{kiis}^H \mathbf{a}_{kiis}|^2}{\sum_{(m,j)\neq(k,i)} \mathrm{E}|\mathbf{h}_{mjis}^H \mathbf{a}_{kiis}|^2 + \mathrm{E}|\tilde{\mathbf{w}}_{is}^H \mathbf{a}_{kiis}|^2}. \qquad (10)$$

The beamforming vector $\mathbf{a}_{kiis}$ exploits the estimated channel (8) which is a function of $\mathbf{v}_{kis}$, $\mathbf{x}_{mj}, \forall m, j$ and $N_c$. Hence, the achievable $\bar{\gamma}_{kis}$ depends on these variables that need to be optimized which is the focus of the next section.

## V. OPTIMIZATION OF $N_c$, $\mathbf{x}_{mj}$ AND $\mathbf{v}_{kis}$

The achievable $\bar{\gamma}_{kis}$ depends on $\mathbf{a}_{kiis}$ where one can utilize different beamforming methods to design $\mathbf{a}_{kiis}$. In a massive MIMO system, simple beamforming methods such as maximum ratio combining (MRC) and zero forcing (ZF) are close to optimal which motivates us to choose $\mathbf{a}_{kiis}$ to be the MRC beamformer in this subsection[4] [4], [6], [10]. That is $\mathbf{a}_{kiis} = \hat{\mathbf{h}}_{kiis} = [\hat{h}_{kii1s}, \hat{h}_{kii2s} \cdots, \hat{h}_{kiiNs}]^T$, where $\hat{h}_{kiins}$ is the estimated channel defined in (8). With this beamforming vector, we will have

$$\mathrm{E}|\mathbf{w}_i^H \mathbf{a}_{kiis}|^2 = \mathbf{v}_{kis}^H \left( \sum_{m=1}^{\tilde{K}} \sum_{j=1}^{N_c} \mathbf{X}_{mj}^* \mathbf{C}_{mji} \mathbf{X}_{mj}^T + \sigma^2 \mathbf{I} \right) \mathbf{v}_{kis}$$

$$\mathrm{E}|\mathbf{h}_{mjis}^H \mathbf{a}_{kiis}|^2 = \mathbf{v}_{kis}^H \left( \mathbf{X}_{mj}^* \left[ \mathbf{C}_{mji} \mathbf{f}_s^* \mathbf{f}_s^T \mathbf{C}_{mji} + \sum_{u=1}^{\tilde{K}} \sum_{v=1,(u,v)\neq(m,j)}^{N_c} \mathbf{C}_{mjuvis} \right] \mathbf{X}_{mj}^T + \sigma^2 \mathrm{tr}\{\mathbf{C}_{mjis}\} \mathbf{I} \right) \mathbf{v}_{kis}$$

with $\mathbf{C}_{mji} = \mathrm{tr}\{\mathbf{R}_{mji}\} \mathbf{P}_{mji}^*$ and $\mathbf{C}_{mjis} = (\mathbf{f}_s^H \mathbf{P}_{mji} \mathbf{f}_s) \mathbf{R}_{mji}^*$, $\mathbf{C}_{mjuvis} = (\mathbf{f}_s^H \mathbf{P}_{uvi} \mathbf{f}_s) \mathrm{tr}\{\mathbf{R}_{uvi}^* \mathbf{R}_{mji}^*\} \mathbf{P}_{mji}^*$ (see Appendix A).

To optimize $\mathbf{v}_{kis}$, $\mathbf{x}_{mj}$ and $N_c$, we examine two problems: In the first problem, we determine the number of cells $N_c$, and feasible $\mathbf{v}_{kis}$ and $\mathbf{x}_{mj}$ such that the multi-cell system is able to mitigate the pilot contamination. In the second problem, we re-optimize $\mathbf{v}_{kis}$ to further maximize $\bar{\gamma}_{kis}$ for fixed $N_c$ and $\mathbf{x}_{mj}, \forall m, j$.

### A. Determination of $N_c$, and Feasible $\mathbf{v}_{kis}$ and $\mathbf{x}_{mj}$

This subsection determines $N_c$, and feasible $\mathbf{v}_{kis}$ and $\mathbf{x}_{mj}$ such that the multicell system is able to mitigate the effect of pilot contamination. Specifically, we determine these variables while ensuring that $\bar{\gamma}_{kis}$ increases as $N$ increases (i.e., $\bar{\gamma}_{kis}$ grows to infinity as $N \to \infty$) in the following Theorem.

*Theorem 1*: For arbitrary channel covariance matrices $\mathbf{P}_{kji}$ and $\mathbf{R}_{kij}$, $\bar{\gamma}_{kis}$ of (10) increases as $N$ increases (i.e., the effect of pilot contamination is mitigated) when $N_c \leq L$, $\mathbf{x}_{mj}, \forall m, j$ are selected such that $\mathbf{Q}_{is}^T, \forall i, s$ are full row rank matrices and $\mathbf{v}_{kis}$ is designed from the solution of

$$\mathbf{D}_{is} \bar{\mathbf{U}}_{is} \mathbf{v}_{kis} = \mathbf{u}_{kiis} \qquad (11)$$

where $\mathbf{Q}_{is} = [\mathbf{q}_{11is}, \mathbf{q}_{21is}, \cdots, \mathbf{q}_{\tilde{K}1is}, \mathbf{q}_{12is}, \mathbf{q}_{22is}, \cdots, \mathbf{q}_{\tilde{K}2is}, \cdots, \mathbf{q}_{1N_cis}, \mathbf{q}_{2N_cis}, \cdots, \mathbf{q}_{\tilde{K}N_cis}]$, $\mathbf{q}_{mjis}^T = \mathbf{f}_s^T \mathbf{C}_{mji} \mathbf{X}_{mj}^T$, $\mathbf{U}_{is}^H \mathbf{D}_{is} \bar{\mathbf{U}}_{is} = \mathrm{SVD}(\mathbf{Q}_{is}^T)$, $\mathbf{U}_{is} \in \mathcal{C}^{N_p \times N_p}$ and $\bar{\mathbf{U}}_{is} \in \mathcal{C}^{N_p \times N_p}$ are unitary matrices, $\mathbf{u}_{kiis}^H$ is the row vector of $\mathbf{U}_{is}^H$ corresponding to $\mathbf{q}_{kiis}^T$ and $\mathbf{D}_{is} \in \mathcal{C}^{N_p \times N_p}$ is a diagonal matrix containing the singular values of $\mathbf{Q}_{is}^T$.

[4]The extension of the results of this section under the uplink and downlink channel ZF, and downlink channel MRC beamformings are discussed in Section VI.

*Proof:* See Appendix B. ∎

For arbitrary covariance matrices $\mathbf{C}_{(.)}$, one approach of satisfying full rank $\mathbf{Q}_{is}^T$ is by selecting $\mathbf{x}_{mj}, \forall m, j$ so that they will be uncorrelated to each other. Furthermore, in order to maintain full rank $\mathbf{Q}_{is}^T$ for all sub-carriers, $\mathbf{x}_{mj}, \forall m, j$ may need to have a "white noise" like sequence. This motivates us to select $\mathbf{x}_{11}$ from noise like deterministic (or random) sequence, and then construct $\mathbf{x}_{mj}, \forall (m, j) \neq (1, 1)$ by shifting the elements of $\mathbf{x}_{11}$ with $N_p$ sized FFT matrix. With such selection, however, we have noticed that the average rate achieved by one sub-carrier differs from that of other subcarriers and this rate imbalance may not be desirable in practice. For this reason, we choose $\mathbf{x}_{11}$ randomly from a given set of sequences where these sequences are known a priori to all BSs and UEs. Upon doing so, we have observed that each sub-carrier of a UE achieves the same average rate.

One can understand from the result of *Theorem 1* that if $N_c > L$, the proposed channel estimation and beamforming approach will achieve "bounded $\bar{\gamma}_{kis}$" even though $N \to \infty$ (i.e., if $N_c > L$, we will experience pilot contamination as will be demonstrated in the simulation section).

### B. Re-optimization of $\mathbf{v}_{kis}$

From the above subsection, we are able to determine $N_c$, and provide feasible $\mathbf{v}_{kis}$ and $\mathbf{x}_{mj}$ while ensuring that the effect of pilot contamination is mitigated. For the given $N_c$, it is also possible to optimize $\mathbf{v}_{kis}$ and $\mathbf{x}_{mj}$ to further increase $\bar{\gamma}_{kis}$. However, jointly optimizing $\mathbf{v}_{kis}$ and $\mathbf{x}_{mj}$ to maximize $\bar{\gamma}_{kis}$ is still complicated. For this reason, we re-optimize $\mathbf{v}_{kis}$ only to maximize $\bar{\gamma}_{kis}$ for fixed $N_c$ and $\mathbf{x}_{mj}$ as follows.

$$\max_{\mathbf{v}_{kis}} \bar{\gamma}_{kis}. \qquad (12)$$

This problem is a Rayleigh quotient optimization problem where generalized eigenvalue solution approach can be applied to get the optimal $\mathbf{v}_{kis}$ [19], [22], [23].

The next issue is to study the relation between the solutions obtained from (11) and (12)? Denote the solution obtained by solving (12) as $\mathbf{v}_{kis}^*$ and its corresponding SINR as $\gamma_{kis}^*$. As this problem is a Rayleigh quotient, $\mathbf{v}_{kis}^*$ is a global optimal solution. Hence, the SINR obtained by any $\mathbf{v}_{kis} \neq \mathbf{v}_{kis}^*$ cannot be higher than that of $\gamma_{kis}^*$ which leads to (after some mathematical steps)

$$\gamma_{kis}^* \geq \bar{\gamma}_{kis} \qquad (13)$$

where
$$\bar{\gamma}_{kis} = \frac{N+a}{b}\bigg|_{\mathbf{v}_{kis}= \text{soln. of (11)}}, \quad (14)$$

$a = \frac{1}{N}\mathbf{v}_{kis}^H(\sum_{u=1}^{\tilde{K}}\sum_{v=1,(u,v)\neq(k,i)}^{N_c}\mathbf{X}_{mj}^*\mathbf{C}_{kiuvis}\mathbf{X}_{mj}^T + \sigma^2\text{tr}\{\mathbf{C}_{kiis}\}\mathbf{I})\mathbf{v}_{kis}$, $b = \frac{1}{N}\mathbf{v}_{kis}^H(\sum_{m=1}^{\tilde{K}}\sum_{j=1}^{N_c}\sigma^2\mathbf{X}_{mj}^*\mathbf{C}_{mji}\mathbf{X}_{mj}^T + \sum_{(m,j)\neq(k,i)}\sum_{(u,v)\neq(m,j)}\mathbf{X}_{mj}^*\mathbf{C}_{mjuvis}\mathbf{X}_{mj}^T + \sigma^2\text{tr}\{\mathbf{C}_{mjs}\}\mathbf{I} + \sigma^4\mathbf{I})\mathbf{v}_{kis}$. As can be seen from this expression, $a$ and $b$ are constant terms which are not dependent on $N$. Hence, for fixed $N_c, L$ and $\tilde{K}$, one can expect that the SINR obtained by utilizing $\mathbf{v}_{kis}$ of (11) and (12) will be almost the same for large $N$. However, the solutions of these two problems may have different SINRs when $N$ is not sufficiently large. This fact has been verified in the simulation section.

We would like to emphasize here that the current paper utilizes pilot transmission without OFDM as shown in (6). However, as clearly seen from this equation, performing pilot transmission without OFDM "alone" would not increase $N_c$. Nevertheless, by performing pilot transmission without OFDM, data transmission with OFDM, carefully selecting $\mathbf{x}_{mj}$, and introducing and optimizing the variable $\mathbf{v}_{kis}$, we are able to increase the number of cells to $N_c = L$. Furthermore, as the proposed design serves $\tilde{K}L = N_p$ UEs in all cells, one can notice that our approach utilizes only one "net" pilot per UE for any $L$ (i.e., a reduction of the utilized pilots per UE by a factor of $L$ compared to the existing approach (4)).

## VI. EXTENSION OF THE PROPOSED APPROACH FOR OTHER BEAMFORMING AND CHANNEL TYPES

The number of cells provided in the above section is derived under the assumption that data transmission takes place in the uplink channel. This section discusses the extension of the proposed design for the downlink channel transmission and the uplink (downlink) data transmission with ZF beamforming.

### A. Extension of the Proposed Approach for Downlink Channel Data Transmission

In this section, we extend the analysis of the above section when data transmission takes place in the downlink channel. For this channel, each BS will perform beamforming to the UEs corresponding to its own cell. When the BS in cell $i$ utilizes the beamforming vectors, the following signal is transmitted by the $i$th BS in sub-carrier $s$

$$\bar{\mathbf{d}}_{is} = \sum_{k=1}^{\tilde{K}}\mathbf{b}_{kiis}d_{kis} \quad (16)$$

where $\mathbf{b}_{kiis} = \lambda_{is}\hat{\mathbf{h}}_{kiis}$ is the precoding vector for the $k$th UE in cell $i$ and sub-carrier $s$, and $\lambda_{is}$ is introduced to maintain the average transmitted power at the $i$th BS $s$th sub-carrier as desired. Thus, in sub-carrier $s$, the $k$th UE in cell $i$ will receive the following signal from all BSs

$$r_{kis}^d = \sum_{j=1}^{N_c}\mathbf{h}_{kijs}^H\bar{\mathbf{d}}_{js} + n_{kis} = \sum_{j=1}^{N_c}\mathbf{h}_{kijs}^H(\sum_{u=1}^{\tilde{K}}\mathbf{b}_{ujjs}d_{ujs}) + \bar{w}_{kis}$$

$$= \mathbf{h}_{kiis}^H\mathbf{b}_{kiis}d_{kis} + \sum_{m=1}^{\tilde{K}}\sum_{j=1,(m,j)\neq(k,i)}^{N_c}\mathbf{h}_{kijs}^H\mathbf{b}_{mjjs}d_{mjs} + \bar{w}_{kis}$$

(17)

where $\bar{w}_{kis}$ is the noise sample of the $k$th UE in the $s$th sub-carrier of cell $i$ which is assumed to be a ZMCSCG random variable with variance $\sigma^2$.

Following the approach of the above section, we will have

$$\mathrm{E}\{\mathbf{h}_{kijs}^H\mathbf{b}_{mjjs}\} = \mathrm{E}\{\mathbf{f}_s^T\bar{\mathbf{H}}_{kij}^*\mathbf{R}_j^T\mathbf{v}_{mjs}\}$$
$$= \mathbf{f}_s^T(\sum_{u=1}^{\tilde{K}}\sum_{v=1}^{N_c}(\bar{\mathbf{H}}_{kij}^*\bar{\mathbf{H}}_{uvj}^T)\mathbf{X}_{uv}^T + \bar{\mathbf{H}}_{kij}^*\bar{\mathbf{W}}_j^T)\mathbf{v}_{mjs}$$
$$= \lambda_{js}\mathbf{f}_s^T\mathbf{C}_{kij}\mathbf{X}_{ki}^T\mathbf{v}_{mjs} \quad (18)$$

By employing the technique of (11), one can select $\mathbf{v}_{mjs}, \forall m, j, s$ ensuring

$$\mathbf{f}_s^T\mathbf{C}_{kii}\mathbf{X}_{ki}^T\mathbf{v}_{kis} = \lambda_{is}N,$$
$$\mathbf{f}_s^T\mathbf{C}_{kij}\mathbf{X}_{ki}^T\mathbf{v}_{mjs} = 0, \ (m,j) \neq (k,i) \quad (19)$$

when $N_c \leq L$. The $\mathbf{v}_{mjs}, \forall m, j, s$ can also be optimized as

$$\max_{\mathbf{v}_{mjs}, \forall m,j,s} \gamma_{kis} \quad (20)$$

where

$$\gamma_{kis} = \frac{\mathrm{E}\{|\mathbf{h}_{kiis}^H\mathbf{b}_{kiis}|^2\}}{\sum_{m=1}^{\tilde{K}}\sum_{j=1,(m,j)\neq(k,i)}^{N_c}\mathrm{E}\{|\mathbf{h}_{kijs}^H\mathbf{b}_{mjjs}|^2\} + \sigma^2}$$

(21)

$$\mathrm{E}\{|\mathbf{h}_{kijs}^H\mathbf{b}_{mjjs}|^2\} =$$
$$\mathrm{E}\{|\mathbf{f}_s^T(\sum_{u=1}^{\tilde{K}}\sum_{v=1}^{N_c}(\bar{\mathbf{H}}_{kij}^*\bar{\mathbf{H}}_{uvj}^T)\mathbf{X}_{uv}^T + \bar{\mathbf{H}}_{kij}^*\bar{\mathbf{W}}_j^T)\mathbf{v}_{mjs}|^2\} \quad (22)$$
$$= \lambda_{js}^2\mathbf{v}_{mjs}^H\bigg(\mathbf{X}_{ki}^*\mathbf{C}_{kij}\mathbf{f}_s^*\mathbf{f}_s^T\mathbf{C}_{kij}\mathbf{X}_{mi}^T + \sum_{u=1}^{\tilde{K}}\sum_{v=1,(u,v)\neq(m,j)}^{N_c}$$
$$\mathbf{X}_{uv}^*\mathbf{C}_{uvis}\mathbf{X}_{uv}^T + \sigma^2\text{tr}\{\mathbf{C}_{kijs}\}\mathbf{I}\bigg)\mathbf{v}_{mjs}.$$

Since the denominator term of $\gamma_{kis}$ incorporates $\mathbf{v}_{mjs}, \forall(m,j) \neq (k,i)$, the convexity of problem (20) is not clear (for instance (20) is not a Rayleigh quotient problem). For this reason, we are not able to get the optimal solution for (20). This phenomena is not surprising since downlink channel problems are often tend to have more complicated mathematical structures than those of the uplink channel [24]. One approach of optimizing (20) will be to employ the solution obtained from (19). By doing so, the achieved $\gamma_{kis}$ is given in (15)

As can be seen from this equation, $\gamma_{kis}$ increases as $N$ increases which is expected to effectively mitigate pilot contamination in multicell systems. One can observe from (21) and (15) that the SINR expressions of the uplink and downlink

$$\gamma_{kis} = \frac{N\lambda_{js}^2 + \lambda_{js}^2 \mathbf{v}_{mjs}^H \left( \sum_{u=1}^{\tilde{K}} \sum_{v=1,(u,v)\neq(m,j)}^{N_c} \lambda_{js} \mathbf{X}_{uv}^* \mathbf{C}_{uvis} \mathbf{X}_{uv}^T + \sigma^2 \text{tr}\{\mathbf{C}_{kijs}\}\mathbf{I} \right) \mathbf{v}_{mjs}}{\sum_{m=1}^{\tilde{K}} \sum_{j=1,(m,j)\neq(k,i)}^{N_c} \lambda_{js}^2 \mathbf{v}_{mjs}^H \left( \sum_{u=1}^{\tilde{K}} \sum_{v=1,(u,v)\neq(m,j)}^{N_c} \mathbf{X}_{uv}^* \mathbf{C}_{uvis} \mathbf{X}_{uv}^T + \sigma^2 \text{tr}\{\mathbf{C}_{kijs}\}\mathbf{I} \right) \mathbf{v}_{mjs} + \sigma^2}. \quad (15)$$

expressions are slightly different (see also [4]). However, these SINR expressions behave similarly (i.e., both the uplink and downlink SINRs will grow unboudedly when $N \to \infty$).

### B. Extension of the Proposed Approach for ZF Beamforming

This subsection discusses the proposed approach by considering that the BS employs ZF beamforming both for uplink and downlink channels. In this regard, we utilize a two step approach as follows. First $\mathbf{V}_{is} = [\mathbf{v}_{1is}, \mathbf{v}_{2is}, \cdots, \mathbf{v}_{\tilde{K}is}]$ in the uplink and downlink channels are optimized jointly using (11) and (19), respectively, to cancel out-of-cell interference. By doing so and very large $N$, the received signal corresponding to the $k$th UE in cell $i$ can be approximated as

$$r_{kis}^u \approx \mathbf{a}_{kiis}^H (\sum_{m=1}^{\tilde{K}} \mathbf{h}_{miis} d_{mis} + \tilde{\mathbf{w}}_{is}), \quad \text{UL} \quad (23)$$

$$r_{kis}^d \approx \mathbf{h}_{kiis}^H \bar{\mathbf{d}}_{is} + \bar{w}_{kis} = \mathbf{h}_{kiis}^H (\sum_{m=1}^{\tilde{K}} \mathbf{b}_{miis} d_{uis}) + \bar{w}_{kis}, \text{ DL}$$

where $(.)^u$ and $(.)^d$ denote uplink and downlink, respectively. Then, $\mathbf{a}_{miis}, \forall m$ and $\mathbf{b}_{miis}, \forall m$ are recomputed by utilizing $\hat{\mathbf{H}}_{is} = [\hat{\mathbf{h}}_{1iis}, \hat{\mathbf{h}}_{2iis}, \cdots, \hat{\mathbf{h}}_{\tilde{K}is}]$ as

$$\mathbf{A}_{is} = \hat{\mathbf{H}}_{is}(\hat{\mathbf{H}}_{is}^H \hat{\mathbf{H}}_{is})^{-1} \Big|_{\mathbf{v}_{kis} = \text{soln. of (11)}}, \quad (24)$$

$$\mathbf{B}_{is} = \bar{\lambda}_{is} \hat{\mathbf{H}}_{is}(\hat{\mathbf{H}}_{is}^H \hat{\mathbf{H}}_{is})^{-1} \Big|_{\mathbf{v}_{kis} = \text{soln. of (19)}} \quad (25)$$

where $\bar{\lambda}_{is}$ is introduced to maintain the total transmitted power at the $i$th BS as desirable, $\mathbf{A}_{is} = [\mathbf{a}_{1iis}, \mathbf{a}_{2iis}, \cdots, \mathbf{a}_{\tilde{K}iis}]$ and $\mathbf{B}_{is} = [\mathbf{b}_{1iis}, \mathbf{b}_{2iis}, \cdots, \mathbf{b}_{\tilde{K}iis}]$. From this discussion, we can understand that the estimated channel matrix $\hat{\mathbf{H}}_{is}$ obtained for the uplink and downlink channel data transmission is not necessarily the same (since the optimal/suboptimal $\mathbf{v}_{kis}$ may vary for the uplink and downlink channels)[5]. On the other hand, $r_{kis}$ of (23) do not have a signal from cell $j, j \neq i$ which confirms that cell $i$ is able to efficiently mitigate the effect of pilot contamination.

### VII. PERFORMANCE ANALYSIS AND DISCUSSIONS

This section analyzes the performance of our design, and provides examples and detailed discussions about the proposed design.

---

[5]One can also utilize the $\mathbf{v}_{kis}$ of (12) and (20) to get better $\gamma_{kis}$ with $\mathbf{A}_{is}$ and $\mathbf{B}_{is}$ especially when $N$ is not sufficiently large.

### A. Performance Analysis

In a practical wireless system, the rate achieved by $\hat{d}_{kis}$ (i.e., $R_{kis}$) which is related to $\bar{\gamma}_{kis}$ has an interest. Furthermore, $R_{kis}$ will be maximum when each of the BSs has perfect CSI of its own UEs. On the other hand, the performances of different beamforming schemes discussed in the aforementioned sections behave similarly when $N$ is very large. This motivates us to examine $R_{kis}$ of the proposed approach and that of perfect CSI for the uplink channel with MRC beamforming scenario. In this regard, we consider the following theorem.

*Theorem 2*: When $N$ is large, $\mathbf{P}_{kji} = g_{kji}\mathbf{I}$ and $\mathbf{R}_{kji} = \mathbf{I}$, the following rates can be achievable.

$$R_{kis}^{pr} = \log_2(1 + \gamma_{kis}^{pr}), \quad R_{kis}^{pe} = \log_2(1 + \gamma_{kis}^{pe}) \quad (29)$$

where $R_{kis}^{pe}(\gamma_{kis}^{pe})$ and $R_{kis}^{pr}(\gamma_{kis}^{pr})$ are the rate (SINR) achieved by perfect CSI scenario and proposed method, respectively, with

$$\gamma_{kis}^{pe} = \frac{NLg_{kii}}{\sum_{m=1}^{\tilde{K}} \sum_{j=1,(m,j)\neq(k,i)}^{N_c} Lg_{kji} + \sigma^2}, \quad (30)$$

$\gamma_{kis}^{pr}$ is expressed in (26) where $\tilde{\mathbf{Z}} = \sum_{(m,j)} g_{mji} \mathbf{X}_{mj}^* \mathbf{X}_{mj}^T + \sigma^2 \mathbf{I}$. Specifically, when $g_{kii} \gtrapprox g_{mji}$ which is the case in practice and $N \to \infty$ (or $N \gg L\tilde{K}N_c$), the proposed approach achieves the SINR $\tilde{\gamma}_{kis}^{pr}$ given in (27). Furthermore, when $\gamma_{kis}^{pe} \gg 1$ and $\gamma_{kis}^{pr} \gg 1$, we will have

$$R_{kis}^{pe} - R_{kis}^{pr} = \log_2 \left( \frac{\gamma_{kis}^{pe}}{\gamma_{kis}^{pr}} \right) \leq \log_2(\eta) = c_0$$

$$R_{kis}^{pe} \approx R_{kis}^{pr} \Big|_{N \to \infty} \quad (31)$$

where $\eta$ is as defined in (28) and the approximation is due to the fact that $c_0$ is independent of $N$.

*Proof:* See Appendix C. ∎

Up to now, we have provided the number of cells that can utilize the same time-frequency resource in a multipath fading channel. Furthermore, the above theorem has been proven for large $N$. Thus, this will raise a question whether the proposed design always achieves better SINR than that of the existing designs for arbitrary $N$ (for example MMSE channel estimation and beamforming designs employing OFDM). Addressing this issue analytically for a general system setup appears to be intractable as the exact answer depends on different parameter settings such as the channel covariance matrices of all UEs. We examine this problem for the Rayleigh fading channel scenario under the MRC beamforming uplink channel data transmission in the following Lemma.

*Lemma 1*: When $\mathbf{P}_{kji} = g_{kji}\mathbf{I}$ and $\mathbf{R}_{kji} = \mathbf{I}$, the proposed design achieves better SINR than that of the existing MMSE based channel estimation and beamforming design when

$$N \geq \frac{N_c \tilde{K}(L\tilde{K} - 1)}{L}. \quad (32)$$

$$\gamma_{kis}^{pr} = \frac{\mathbf{v}_{kis}^H \left( \mathbf{X}_{ki}^* \left[ N g_{kii}^2 \mathbf{f}_s^* \mathbf{f}_s^T + \sum_{(u,v) \neq (k,i)} L g_{kii} g_{uvi} \mathbf{I} \right] \mathbf{X}_{ki}^T + L\sigma^2 g_{kii} \mathbf{I} \right) \mathbf{v}_{kis}}{\mathbf{v}_{kis}^H \left( \sum_{(m,j) \neq (k,i)} \left( \mathbf{X}_{mj}^* \left[ N g_{mji}^2 \mathbf{f}_s^* \mathbf{f}_s^T + \sum_{(u,v) \neq (m,j)} L g_{mji} g_{uvi} \mathbf{I} \right] \mathbf{X}_{mj}^T + L\sigma^2 g_{mji} \mathbf{I} \right) + \sigma^2 \tilde{\mathbf{Z}} \right) \mathbf{v}_{kis}} \quad (26)$$

$$\tilde{\gamma}_{kis}^{pr} = \left. \frac{N g_{kii}^2 + \mathbf{v}_{kis}^H \left( \sum_{(u,v) \neq (k,i)} L g_{kii} g_{uvi} \mathbf{X}_{ki}^* \mathbf{X}_{ki}^T + L\sigma^2 g_{kii} \mathbf{I} \right) \mathbf{v}_{kis}}{\mathbf{v}_{kis}^H \left( \sum_{(m,j) \neq (k,i)} \left[ \sum_{(u,v) \neq (m,j)} L g_{mji} g_{uvi} \mathbf{X}_{mj}^* \mathbf{X}_{mj}^T + L\sigma^2 g_{mji} \mathbf{I} \right] + \sigma^2 \tilde{\mathbf{Z}} \right) \mathbf{v}_{kis}} \right|_{\mathbf{v}_{kis}= \text{soln. of } (11)} \quad (27)$$

$$\eta = \left. \frac{L \mathbf{v}_{kis}^H \left( \sum_{(m,j) \neq (k,i)} \left[ \sum_{(u,v) \neq (m,j)} L g_{mji} g_{uvi} \mathbf{X}_{mj}^* \mathbf{X}_{mj}^T + L\sigma^2 g_{mji} \mathbf{I} \right] + \sigma^2 \tilde{\mathbf{Z}} \right) \mathbf{v}_{kis}}{g_{kii} (\sum_{m=1}^{\tilde{K}} \sum_{j=1,(m,j) \neq (k,i)}^{N_c} L g_{kji} + \sigma^2)} \right|_{\mathbf{v}_{kis}= \text{soln. of } (11)} \quad (28)$$

If $N_c = L$, we will have $N \gtrsim \tilde{K}^2 L$.

*Proof:* See Appendix D. ∎

We would like to point out here that the proof of *Lemma 1* is derived with several approximations which are not necessarily tight. And one may get more precise value of $N$ by utilizing tight approximations. Nevertheless, the result of this lemma fits well with that obtained from the simulation (see Section VIII for the details).

### B. Example and Discussions

From the above Lemma, we can realize that $N$ is related to $\tilde{K}$ and $L$. For better insight into the proposed method, we provide numerical examples by employing the parameters used in the LTE network. For this network, different possible bandwidths are utilized (see [17], [18] for more details). Here we consider the case where $B = 2.5$MHz with $T_d = 4.69\mu s$, coherence time $T_c = 0.5$ms, $T_p = \frac{T_c}{2}$ (i.e., the pilot duration that maximizes the overall throughput [4]) and $T_s = \frac{66.7}{256} = 0.2605\mu s$ [4], [18]. With these parameter settings, we get

$$\tilde{K} = \frac{T_p}{T_d} \approx 53, \quad N_c = L = \frac{T_d}{T_s} \approx 18. \quad (33)$$

One can notice from (33) that our approach can serve $L\tilde{K} \approx 954$ UEs while mitigating pilot contamination. This is similar to scaling the number of UEs using the same resource by $L$. In fact, $T_d$ is constant for a particular environment and carrier frequency, and hence $L$ increases as the bandwidth increases [16]. Thus, doubling $B$ will double $L$ which consequently doubles the number of cells $N_c$. However, according to the result of Lemma 1, the proposed approach achieves better performance than that of the existing ones when $N \approx \tilde{K}^2 L = 50526$ for the above parameter settings which is not realistic. This will raise a concern on the advantage of the proposed design for a realistic value of $N$ (for instance $N \leq 256$).

As we can see from (33), $\tilde{K}$ and $L$ can be modified by controlling $T_p$ and $B$, respectively. Thus, for arbitrary $N$, it is always possible to choose $\tilde{K}$ and $L$ appropriately such that the proposed design can achieve better performance than that of the existing MMSE design. When $\tilde{K} = L = 4$ (see Figs. 1 and 5 in Section VIII), we have observed from the simulation results that the proposed design achieves significantly better performance than the existing designs for $N \geq 64$. Having said this, however, how to jointly select $\tilde{K}$ and $L$ ensuring the optimal throughput is beyond the scope of the current paper and is left for future research.

We would like to mention here that the analysis and discussions of this section is provided for the MRC based uplink data transmission under the simplified assumption where $\mathbf{P}_{kji} = g_{kji}\mathbf{I}$ and $\mathbf{R}_{kji} = \mathbf{I}$ (i.e., Rayleigh fading channel environment). Extending the analysis of this section to other system setup and channel conditions is omitted for conciseness.

### C. Practical Issues

The current paper provides two approaches to obtain $\mathbf{v}_{kis}$ (i.e., with (11) and (12)). The solution of (12) assumes that the $i$th BS has perfect knowledge of the covariance matrices (i.e., $\mathbf{P}_{mji}$ and $\mathbf{R}_{mji}$) associated with the CSI between the UEs in all cells and the $i$th BS. Furthermore, since $\mathbf{P}_{mji}$ has $L$ coefficients only which are typically related to the distance dependent path loss, it can be estimated reliably with manageable complexity. However, assuming perfect $\mathbf{R}_{mji}$ may not be practically appealing for two main reasons: The first concern is that these covariance matrices are estimated from the long term statistics which often exhibit estimation error. The other and most important concern is that the $i$th BS may need to estimate $\mathbf{R}_{mji}, \forall m, j$ which creates considerable signaling overhead especially when $N(L)$ is very large. In fact, the solution of (11) does not require the spatial covariance matrix which helps reduce the design complexity as the dimension of $\mathbf{R}_{mji}$ scales with $N$. However, this complexity reduction is at the cost of performance degradation especially when $N$ is not very large. This shows that obtaining the optimal $\mathbf{v}_{kis}$ of (12) with imperfect, unknown or local (i.e., each BS exploits the $\mathbf{R}_{mji}$ of its own UEs only) spatial channel covariance matrix is still an open research topic which is left for future work[6].

The proposed joint channel estimation and beamforming approache is summarized in **Algorithm I**.

**Algorithm I: Summary of the Proposed Channel Estimation and Beamforming Design**

1) Initializations:
    - Fix $B$, $T_p$, $T_o$, $T_s$, $T_d$, $T_u$, $L$ and $\tilde{K}$.

---

[6] We would like to point out here that such a problem can potentially be addressed by leveraging the techniques used in the robust beamforming designs (see for example [25]–[27]).

- Set $N_c \leq L$ as desirable (large $N_c$ to serve more UEs) and their corresponding $\mathbf{P}_{kji}$ and $\mathbf{R}_{kji}$.
- Select the pilots (7) (fixed or randomly selected).

2) **If Uplink Channel MRC Beamforming is desired**
   **Channel Estimation**:
   - Compute the estimate of $h_{kiins}$ (i.e., $\hat{h}_{kiins}$) using (8) where $\mathbf{v}_{kis}$ is determined from (11) or from the solution of (12). Note that the complexity of solving the latter problem is almost the same as the former one (i.e., the complexity of SVD).
   
   **Beamforming**
   - Using the $h_{kiins}$ of the above step, use the MRC beamforming vector of the $k$th UE in cell $i$ (i.e., $\mathbf{a}_{kiis} = \hat{\mathbf{h}}_{kiis} = [\hat{h}_{kii1s}, \hat{h}_{kii2s}, \cdots, \hat{h}_{kiiNs}]^T$ to decode $d_{kiis}$.

3) **If Downlink Channel MRT Beamforming is desired**
   **Channel Estimation**:
   - Compute the estimate of $h_{kiins}$ (i.e., $\hat{h}_{kiins}$) using (8) where $\mathbf{v}_{kis}$ is determined from (19).
   
   **Beamforming**:
   - Using the $h_{kiins}$ of the above step, use the MRT beamforming vector of the $k$th UE in cell $i$ (i.e., $\mathbf{b}_{kiis} = \hat{\mathbf{h}}_{kiis} = [\hat{h}_{kii1s}, \hat{h}_{kii2s}, \cdots, \hat{h}_{kiiNs}]^T$ to precode $d_{kiis}$).

4) **If Uplink Channel ZF Beamforming is desired**
   **Channel Estimation**:
   - Compute the estimate of $h_{kiins}$ (i.e., $\hat{h}_{kiins}$) using (8) where $\mathbf{v}_{kis}$ is determined from (11).
   
   **Beamforming**: Using the $h_{kiins}$ of the above step, use the ZF beamforming matrix of the UEs in cell $i$ (i.e., $\mathbf{A}_{is}$ to decode $d_{kiis}, \forall k$).

5) **If Downlink Channel ZF Beamforming is desired**
   **Channel Estimation**:
   - Compute the estimate of $h_{kiins}$ (i.e., $\hat{h}_{kiins}$) using (8) where $\mathbf{v}_{kis}$ is determined from (19).
   
   **Beamforming**:
   - Using the $h_{kiins}$ of the above step, use the ZF beamforming matrix of the UEs in cell $i$ (i.e., $\mathbf{B}_{is}$ to precode $d_{kiis}, \forall k$).

## VIII. SIMULATION RESULTS

This section presents simulation results by considering a multicell system with $N_c$ cells where each cell serves $\tilde{K} = 4$ UEs. In this regard, we set $L = 4$, $N_p = 16$, $M = 64$, $N = 2^{N_0}$ and $k = i = 1$ (i.e., the target user and cell) where $N_0$ is an integer which is selected according to the desired $N$. We examine the average achievable rate of the UE's subcarrier (i.e., $R_{kis}$). The signal to noise ratio (SNR) is defined as $\gamma = \frac{\mathrm{E}\{|d_{ki}|^2\}}{\tilde{\sigma}^2}, \forall k, i, \tilde{\sigma}^2 = \sigma^2$. For conciseness, the simulation results are presented for the case where data transmission takes place in the uplink channel with MRC beamforming approach. All results are plotted by averaging 10000 channel realizations.

### A. Comparison of the Proposed and Existing Approaches

This subsection compares the performance of the proposed approach with those of the existing ones when $\mathbf{P}_{kji} = g_{kji}\mathbf{I}$ and $\mathbf{R}_{kji} = \mathbf{I}$ with $g_{k1i} = 1, g_{k2i} = 0.9, g_{k3i} = 0.6, g_{k4i} =$ $0.7, \forall k$. In the proposed approach, first we choose $\mathbf{x}_{11}$ from a randomly selected unit energy quadrature phase shift keying (QPSK) signal. Then, we compute the remaining pilots just by shifting $\mathbf{x}_{11}$ appropriately with equally spaced conventional (fractional) discrete Fourier transform (DFT) vectors. For the existing approaches, we design $\mathbf{x}_{ki}, \forall i, k$ in two methods: the first method uses orthogonal pilots like in the proposed approach and the second method applies full pilot reuse (i.e., $\mathbf{x}_{ki} = \mathbf{x}_{kj}, \forall i, j, k$).

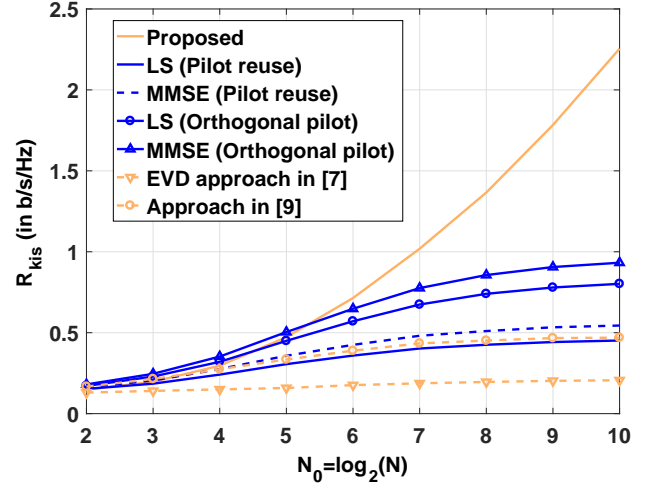

Fig. 1. Comparison of the rates achieved by the proposed and existing approaches.

For fair comparison, we have utilized $\nu = 1$ (to resolve the multiplicative factor ambiguity) and binary phase shift keying transmitted signal of size 50 for the EDV approach of [7], and a full pilot reuse for the approach in [9]. Fig. 1 shows the achieved rates of the proposed and existing approaches when $\gamma = 0$dB, $s = 20$ and $N_c = L$. For the proposed approach we have employed $\mathbf{v}_{kis}$ obtained using (12). From this figure, we can observe that utilizing orthogonal pilots slightly improve the performances of the MMSE and LS methods. Furthermore, the approaches of [7] and [9] also behave similar to the MMSE and LS approaches. However, the rate achieved by the proposed approach increases progressively with $N$. And, our approach achieves significantly higher $R_{kis}$ compared to those of the existing approaches especially in a massive MIMO regime which is desirable.

### B. Validation of Theoretical Results

This subsection validates the theoretical results provided in Sections V and VII when the temporal and spatial covariance matrices are set to $\mathbf{P}_{kji} = \mathbf{I}$ and $\mathbf{R}_{kji} = g_{kji}\mathbf{I}$ with $g_{k1i} = 1, g_{k2i} = 0.9, g_{k3i} = 0.6, g_{k4i} = 0.7, \forall k$. According to *Theorem 1*, if $N_c > 4$, increasing $N$ indefinitely does not help to improve $\bar{\gamma}_{kis}$. Also, the rate achieved by utilizing the solutions of (11) and (12) will be close to each other when $N$ is very large and $c_0$ of (31) is derived for $N \rightarrow \infty$. These motivate us to perform simulation for this subsection by considering very large $N$.

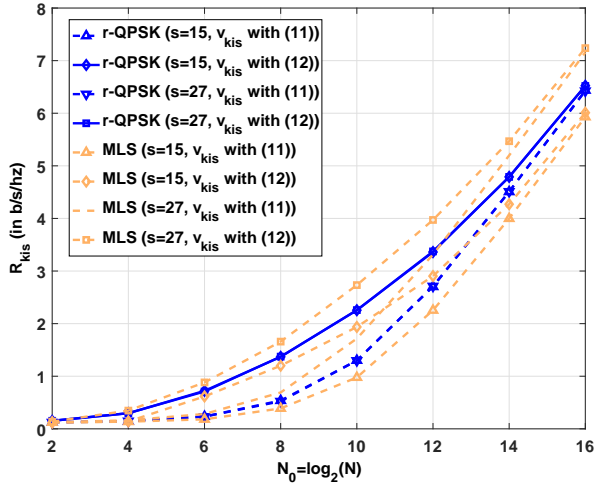

Fig. 2. Comparison of the rates achieved by fixed and random pilots.

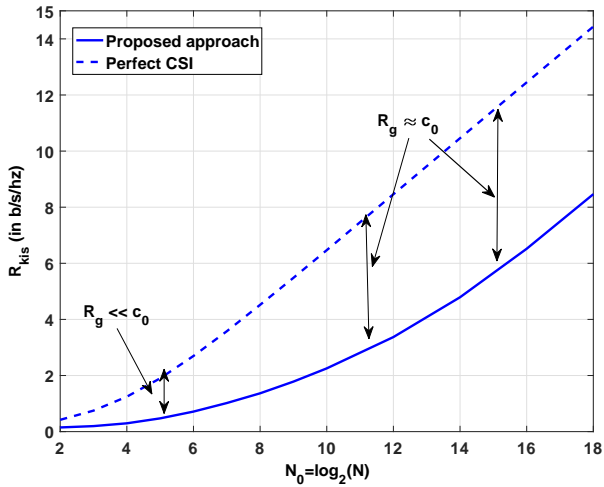

Fig. 3. The rates achieved by the proposed and perfect CSI scenarios.

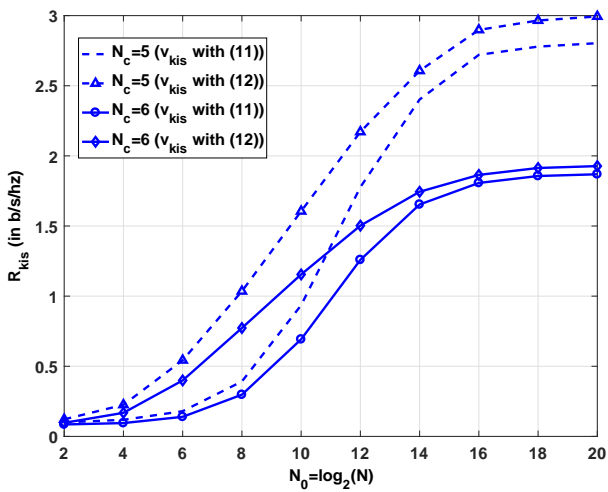

Fig. 4. The rate achieved by the proposed approach for different values of $N_c$.

*1) Effect of Pilots:* This simulation examines the effect of pilots on the performance of the proposed algorithm. The pilot samples are designed by two approaches: The first approach utilizes $\mathbf{x}_{11} = N_p^{-1/2}(\mathbf{z}_1 + \sqrt{-1}\mathbf{z}_2)$ where $[\mathbf{z}_1\ \mathbf{z}_2]$ are the last $2N_p$ samples generated from the "Maximum Length Sequence" $\pm 1$ bits of size 64 (i.e., fixed pilot assignment approach). The second approach designs the pilots as in Section VIII-A (i.e., the pilot symbols are selected randomly but are known a priori to each UE and BS). We examine the rate $R_{kis}$ (i.e., rate per user per sub-carrier) when $\gamma = 0$dB and $s = \{15, 27\}$ as shown in Fig. 2. As can be seen from this figure, in both sub-carriers, $R_{kis}$ increases with $N$ by utilizing either fixed or randomly generated pilots.

On the other hand, the average $R_{kis}$ may not be necessarily the same for all sub-carriers when fixed pilot assignment approach is utilized. However, the rate of these sub-carriers are the same when pilots are selected randomly which is expected (i.e., for given design approach of $\mathbf{v}_{kis}$). This simulation also exploits the fact that the rates achieved by employing the solutions of (11) and (12) are closer to each other when $N$ is very large. However, for small to medium $N$, higher $R_{kis}$ is obtained when $\mathbf{v}_{kis}$ is optimized using (12) which agrees with the theory. In the following simulation, we utilize the $\mathbf{v}_{kis}$ designed by (12) with random pilot assignment approach.

*2) Verification of $c_0$:* This simulation verifies the bound derived in (31) while utilizing $\gamma = 0$dB and $s = 20$ as shown in Fig. 3. From this figure, one can observe that the gap between the rate achieved by the proposed and that of perfect CSI scenario $R_g$ becomes constant ($R_g = c_0 \approx 6$b/s/hz) after $N \approx 2^{16}$ which fits with the theory. On the other hand, $R_g$ decreases as $N$ decreases which is desirable in practice. For instance, if we deploy $N = 16$ antennas, the rate loss is $R_g < 1$b/s/hz. We would like to mention here that, in a massive MIMO regime, we have observed almost the same behavior as in this figure for $\gamma = 20$dB. This is simply because the effect of noise vanishes for sufficiently large $N$ [4], [9].

*3) Effect of $N_c$:* As discussed in Section V, the proposed design experiences pilot contamination if $N_c > L = 4$. This simulation validates this claim. To this end, we set $s = 32$, $N_c = \{5, 6\}$, the channel gains of the UEs corresponding to the first four cells are the same as described in the first paragraph of this section, and the UEs of the fifth and sixth cells have channel gains $g_{k51} = 0.6, g_{k61} = 0.75, \forall k$. Fig. 4 shows the $R_{kis}$ achieved for these settings. As can be seen from this figure and Fig. 3, $R_{kis}$ drops quickly as we increase $N_c$ from 4 to $\{5, 6\}$. Furthermore, in both $N_c = \{5, 6\}$ cases, $R_{kis}$ is not increasing further after approximately $N = 2^{16}$ (i.e., bounded rate) which is in line with the theory provided in *Theorem 1*.

## C. Effect of Channel Correlation

This subsection compares the performance of the proposed and existing designs under the assumption of correlated $\bar{\mathbf{H}}_{kji}$. Towards this end, we assume that $\mathbf{P}_{kji} = g_{kji}\mathbf{I}$ where $g_{kji}$ is like in the above subsection but $\mathbf{R}_{kji}$ is taken from the well known exponential correlation model (i.e., the $(n,m)$th element of $\mathbf{R}_{kji}$ is expressed as $\rho_{kji}^{|n-m|}$ where $0 \leq \rho_{kji} < 1$).

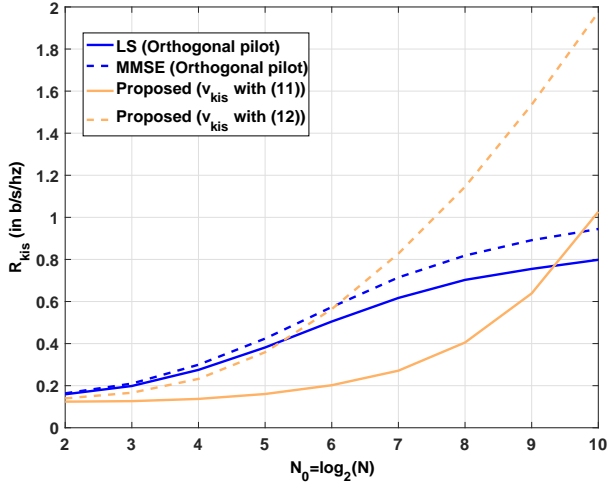

Fig. 5. Comparison of the proposed and existing designs for correlated channels.

We have employed this very simple correlation model for the following reasons. First, the exponential model is physically reasonable in a way that the correlation between two transmit antennas decreases as the distance between them increases [28] (see also Fig. 2 in [29]). Second, this model has been widely used for an urban area communications where traffic is usually congested [30]. For the simulation of this section, we employ $\{\rho_{k1i} = 0.5, \rho_{k1i} = 0.55, \rho_{k1i} = 0.45, \rho_{k1i} = 0.48\}, \forall k$. Fig. 5 shows the sum rate achieved by the proposed and existing MMSE and LS algorithms. As can be seen from this figure, the proposed algorithm achieves better performance than that of the existing ones. From this figure and Fig. 1, we can also notice that the improvement starts at around $N = \tilde{K}^2 N_c \approx 64$. This demonstrates that the results of *Lemma 1* is also valid both for correlated and uncorrelated channel cases.

In [5] and [31], it has been shown that when each BS is equipped with the uniform linear arrays (ULAs) and UEs have non-overlapping AOAs in $[0, \pi]$, the effect of pilot contamination eliminates asymptotically (see (24) and (26) of [5]). In other words, for these AOA values, the rate achieved by LMMSE approach increases as $N$ increases. In the next simulation, we compare the performance of the proposed design with the LS and LMMSE approaches for $\tilde{K} N_c$ non-overlapping AOAs where each UE has AOA's width of $\frac{2}{3} \frac{\pi}{2 \tilde{K} N_c}$ as shown in Fig. 6 when the number of paths is selected as 4 and 8 (i.e., $P = [4, 8]$ in (24) of [5]). From this figure, one can observe that the rate achieved by both the proposed and existing designs are very close to each other and increase as $N$ increases for both $P$ values which is expected.

## IX. CONCLUSIONS

This paper proposes a novel joint channel estimation and beamforming approach for multicell wideband massive MIMO systems experiencing frequency selective channels. With the proposed approach, we determine $N_c$ utilizing the same time-frequency resource while mitigating the effect of pilot contamination for arbitrary channel covariance matrix of each UE. In particular, when the channel has $L$ multipath taps, it is shown that $N_c \leq L$ cells can reliably estimate the channels of their UEs and perform beamforming while efficiently mitigating the effect of pilot contamination. All the theoretical results are demonstrated by using extensive numerical simulations both for correlated and uncorrelated channels. The proposed joint channel estimation and beamforming approach is linear, simple to implement and significantly outperforms the existing approaches.

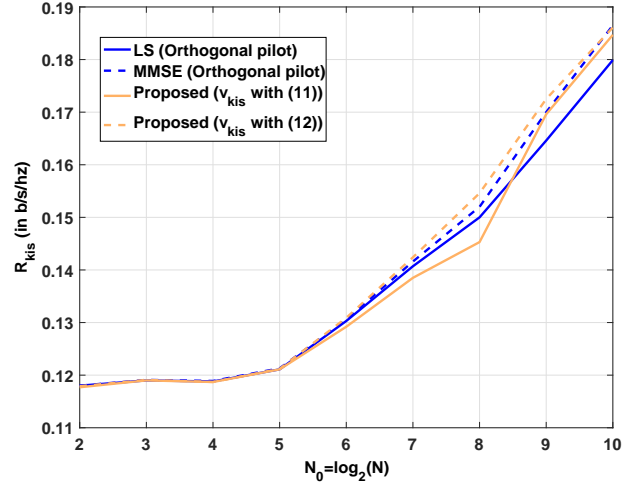

(a)

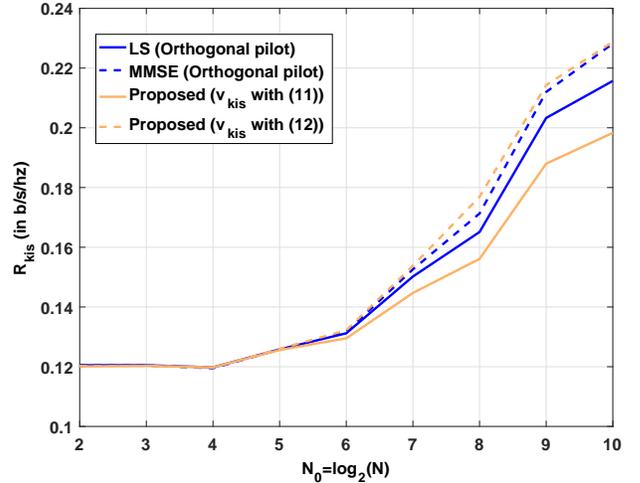

(b)

Fig. 6. Comparison of the proposed and Existing designs for ULA channels with phases as in in (24) of [5]: (a) $P = 4$. (b) $P = 8$.

## APPENDIX A

COMPUTATION OF $\mathrm{E}\{|\mathbf{h}_{mjis}^H \mathbf{a}_{kiis}|^2\}$ AND $\mathrm{E}\{|\mathbf{w}_i^H \mathbf{a}_{kiis}|^2\}$

From (8) and (9), one can obtain

$$\mathrm{E}\{|\mathbf{h}_{mjis}^H \mathbf{a}_{kiis}|^2\} = \quad (34)$$

$$\mathrm{E}\{|\mathbf{f}_s^T (\sum_{u=1}^{\tilde{K}} \sum_{v=1}^{N_c} (\bar{\mathbf{H}}_{mji}^* \bar{\mathbf{H}}_{uvi}^T) \mathbf{X}_{mj} + \bar{\mathbf{H}}_{mji}^* \bar{\mathbf{W}}_i^T) \mathbf{v}_{kis}|^2\}$$

$$=\mathrm{E}\{\mathbf{v}_{kis}^H \bigg(\mathbf{X}_{mj}^*[\bar{\mathbf{H}}_{mji}^* \bar{\mathbf{H}}_{mji}^T \mathbf{f}_s^* \mathbf{f}_s^T \bar{\mathbf{H}}_{mji}^* \bar{\mathbf{H}}_{mji}^T +$$

$$\sum_{u=1}^{\tilde{K}} \sum_{v=1,(u,v)\neq(m,j)}^{N_c} \bar{\mathbf{H}}_{mji}^* \bar{\mathbf{H}}_{uvi}^T \mathbf{f}_s^* \mathbf{f}_s^T \bar{\mathbf{H}}_{uvi}^* \bar{\mathbf{H}}_{mji}^T] \mathbf{X}_{mj}^T +$$

$$\bar{\mathbf{W}}_i^* \bar{\mathbf{H}}_{mji}^T \mathbf{f}_s^* \mathbf{f}_s^T \bar{\mathbf{H}}_{mji}^* \bar{\mathbf{W}}_i^T \bigg) \mathbf{v}_{kis}\}$$

$$=\mathbf{v}_{kis}^H \bigg(\mathbf{X}_{mj}^*[\mathbf{C}_{mji} \mathbf{f}_s^* \mathbf{f}_s^T \mathbf{C}_{mji} + \sum_{u=1}^{\tilde{K}} \sum_{v=1,(u,v)\neq(m,j)}^{N_c} \mathbf{C}_{mjuvis}]$$

$$\mathbf{X}_{mj}^T + \sigma^2 \mathrm{tr}\{\mathbf{C}_{mjis}\} \mathbf{I}\bigg) \mathbf{v}_{kis}$$

$$\mathrm{E}\{|\mathbf{w}_i^H \mathbf{a}_{kiis}|^2\} = \sigma^2 \mathrm{E}\{|\mathbf{a}_{kiis}|^2\} = \sigma^2 \mathrm{E}\{\mathbf{v}_{kis}^H \mathbf{R}_i^* \mathbf{R}_i^T \mathbf{v}_{kis}\}$$

$$= \sigma^2 \mathbf{v}_{kis}^H (\sum_{m=1}^{\tilde{K}} \sum_{j=1}^{N_c} \mathbf{X}_{mj}^* \mathbf{C}_{mji} \mathbf{X}_{mj}^T + \sigma^2 N \mathbf{I}) \mathbf{v}_{kis} \quad (35)$$

where we have defined $\mathbf{C}_{mji} = \mathrm{E}\{\bar{\mathbf{H}}_{mji}^* \bar{\mathbf{H}}_{mji}^T\} = \mathrm{tr}\{\mathbf{R}_{mji}\} \mathbf{P}_{mji}^* = N \mathbf{P}_{mji}^*$, and $\mathbf{C}_{mjis} = \mathrm{E}\{\bar{\mathbf{H}}_{mji}^T \mathbf{f}_s^* \mathbf{f}_s^T \bar{\mathbf{H}}_{mji}^*\} = (\mathbf{f}_s^T \mathbf{P}_{mji} \mathbf{f}_s) \mathbf{R}_{mji}^*$, $\mathbf{C}_{mjuvis} = \mathrm{E}\{\bar{\mathbf{H}}_{mji}^* \bar{\mathbf{H}}_{uvi}^T \mathbf{f}_s^* \mathbf{f}_s^T \bar{\mathbf{H}}_{uvi}^* \bar{\mathbf{H}}_{mji}^T\} = (\mathbf{f}_s^H \mathbf{P}_{uvi} \mathbf{f}_s) \mathrm{tr}\{\mathbf{R}_{uvi}^* \mathbf{R}_{mji}^T\} \mathbf{P}_{mji}^*$. The terms related to $\mathbf{P}_{(.)}$ and $\mathbf{R}_{(.)}$ are obtained by exploiting the fact that $\mathrm{tr}\{\mathbf{R}_{mji}\} = N$ and $\mathrm{E}\{\mathbf{X}^H \mathbf{A} \mathbf{X}\} = \mathrm{tr}\{\mathbf{A}\} \mathbf{I}$ when each entry of $\mathbf{X}$ is an i.i.d ZMCSCG random variable with unit variance [32].

## APPENDIX B

PROOF OF *Theorem 1*

This appendix provides the proof of *Theorem 1*. The goal of this theorem is to determine the number of cells $N_c$ such that the SINR $\bar{\gamma}_{ki}$ increases as $N$ increases. To this end, let us have a closer look at the $\bar{\gamma}_{ki}$ provided in Appendix A. As can be seen from this appendix, $\bar{\gamma}_{ki}$ will be bounded when both the numerator and denominator terms of $\bar{\gamma}_{ki}$ scales with the same factor. In $\mathrm{E}\{|\mathbf{h}_{mjis}^H \mathbf{a}_{kiis}|^2\}$, for instance, the term $\mathbf{v}_{kis}^H \mathbf{X}_{mj}^* \mathbf{C}_{mji} \mathbf{f}_s^* \mathbf{f}_s^T \mathbf{C}_{mji} \mathbf{X}_{mj}^T \mathbf{v}_{kis}$ scales roughly with $N^2$ whereas, $\mathbf{v}_{kis}^H (\mathbf{X}_{mj}^* \sum_{u=1}^{\tilde{K}} \sum_{v=1,(u,v)\neq(m,j)}^{N_c} \mathbf{C}_{mjuvis} \mathbf{X}_{mj}^T + \sigma^2 \mathrm{tr}\{\mathbf{C}_{mjis}\} \mathbf{I}) \mathbf{v}_{kis}$ scales with $N$. This can easily be verified for i.i.d channels (i.e., $\mathbf{R}_{mij} = \mathbf{I}$ and $\mathbf{P}_{mij} = \mathbf{I}$). Hence, one can reexpress $\bar{\gamma}_{ki}$ as

$$\bar{\gamma}_{ki} = \frac{\frac{1}{N} \mathbf{v}_{kis}^H \mathbf{X}_{ki}^* \mathbf{C}_{kii} \mathbf{f}_s^* \mathbf{f}_s^T \mathbf{C}_{kii} \mathbf{X}_{ki}^T \mathbf{v}_{kis} + \beta_0}{\frac{1}{N} \sum_{(m,j)\neq(k,i)} \mathbf{v}_{kis}^H \mathbf{X}_{mj}^* \mathbf{C}_{mji} \mathbf{f}_s^* \mathbf{f}_s^T \mathbf{C}_{mji} \mathbf{X}_{mj}^T \mathbf{v}_{kis} + \beta_1} \quad (36)$$

where $\beta_0$ and $\beta_1$ are the terms that does not depend on $N$.

Thus, one approach of maintaining the property that $\bar{\gamma}_{ki}$ increases as N increases is by designing $\mathbf{v}_{kis}$ such that

$$\max_{\mathbf{v}_{kis}} \mathbf{v}_{kis}^H \bigg(\mathbf{X}_{ki}^* \mathbf{C}_{kii} \mathbf{f}_s^* \mathbf{f}_s^T \mathbf{C}_{kii} \mathbf{X}_{ki}^T\bigg) \mathbf{v}_{kis} \quad (37)$$

$$\mathrm{s.t} \sum_{(m,j)\neq(k,i)} \mathbf{v}_{kis}^H \bigg(\mathbf{X}_{mj}^* \mathbf{C}_{mji} \mathbf{f}_s^* \mathbf{f}_s^T \mathbf{C}_{mji} \mathbf{X}_{mj}^T\bigg) \mathbf{v}_{kis} = 0.$$

One can determine the number of cells $N_c$ by examining this problem. However, the exact solution of this problem depends on the covariance information $\mathbf{C}_{(.)}$. And hence the number of cells could vary from one covariance matrix to another. However, the goal of this appendix is to determine $N_c$ which is valid for all $\mathbf{C}_{(.)}$. In this regard, we have rewritten the above problem (by tightening the constraint) as

$$\max_{\mathbf{v}_{kis}} |\mathbf{f}_s^T \mathbf{C}_{kii} \mathbf{X}_{ki}^T \mathbf{v}_{kis}|$$

$$\mathrm{s.t.} \mathbf{f}_s^T \mathbf{C}_{mji} \mathbf{X}_{mj}^T \mathbf{v}_{kis} = 0, \quad \forall (m,j) \neq (k,i). \quad (38)$$

It can be clearly seen that the solution of (38) is also feasible to (37). Now by defining $\mathbf{q}_{mjis}^T = N^{-1} \mathbf{f}_s^T \mathbf{C}_{mji} \mathbf{X}_{mj}^T$, we can rewrite problem (38) as

$$\max_{\mathbf{X}_{uv}, \forall u,v, \mathbf{v}_{kis}} N|\mathbf{q}_{kiis}^T \mathbf{v}_{kis}|,$$

$$\mathrm{s.t.} \mathbf{Q}_{kis}^T \mathbf{v}_{kis} = \mathbf{0} \quad (39)$$

where $\mathbf{Q}_{kis} = \mathbf{Q}_{is}$ without the column $\mathbf{q}_{kiis}$ and $\mathbf{Q}_{is} = [\mathbf{q}_{11is}, \mathbf{q}_{21is}, \cdots, \mathbf{q}_{\tilde{K}1is}, \mathbf{q}_{12is}, \mathbf{q}_{22is}, \cdots, \mathbf{q}_{\tilde{K}2is}, \cdots, \mathbf{q}_{1N_cis}, \mathbf{q}_{2N_cis}, \cdots, \mathbf{q}_{\tilde{K}N_cis}]$.

From (39), one can reveal that $\mathbf{v}_{kis} \neq 0$ ensuring $\mathbf{Q}_{kis}^T \mathbf{v}_{kis} = \mathbf{0}$ always exists for any $\mathbf{C}_{mji}$ and $\mathbf{X}_{mj}, \forall m, j, i$ if $N_c \leq L$. This is due to the fact that in such a case $\mathbf{Q}_{kis}$ will be rank deficient when $N_c \leq L$. *Therefore, at least $L$ cells can utilize the same time-frequency resource while effectively mitigating the effect of pilot contamination*[7].

Once $N_c$ is determined, the next issue will be to optimize $\mathbf{v}_{kis}$ and $\mathbf{X}_{uv}, \forall u, v$ for fixed $N_c$. This optimization problem can be formulated as

$$\max_{\mathbf{v}_{kis}, \mathbf{X}_{uv}, \forall u,v} N|\mathbf{q}_{kiis}^T \mathbf{v}_{kis}|,$$

$$\mathrm{s.t.} \mathbf{Q}_{kis}^T \mathbf{v}_{kis} = \mathbf{0}. \quad (40)$$

Still jointly optimizing $\mathbf{X}_{uv}, \forall u, v$ and $\mathbf{v}_{kis}, \forall k, i, s$ of (40) is very difficult as the above problem is non-convex. For fixed $\mathbf{X}_{uv}, \forall u, v$, however, the optimal $\mathbf{v}_{kis}$ will lie on the right null space of $\mathbf{Q}_{kis}^T$. To better exploit the structure of the solution of (40), we compute the singular value decomposition (SVD) of $\mathbf{Q}_{is}^T$ as

$$\mathbf{Q}_{is}^T = \mathbf{U}_{is}^H \mathbf{D}_{is} \bar{\mathbf{U}}_{is}$$

where $\mathbf{U}_{is} \in \mathcal{C}^{N_p \times N_p}$ and $\bar{\mathbf{U}}_{is} \in \mathcal{C}^{N_p \times N_p}$ are unitary matrices, and $\mathbf{D}_{is} \in \mathcal{C}^{N_p \times N_p}$ a diagonal matrix containing the singular values of $\mathbf{Q}_{is}^T$.

---

[7]Note that by exploiting the structures of the covariance matrix information of $\bar{\mathbf{H}}_{mji}$ (i.e., $\mathbf{C}_{mji}$), the number of cells can be increased further. However, as the focus of the current paper is to determine $N_c$ for general covariance matrix, special cases are not discussed here.

It follows that $\mathbf{q}_{kiis}^T = \mathbf{u}_{kiis}^H \mathbf{D}_{is} \bar{\mathbf{U}}_{is}$ where $\mathbf{u}_{kiis}^H$ is the row vector of $\mathbf{U}_{is}^H$ corresponding to $\mathbf{q}_{kiis}^T$. From this decomposition, one can observe that the optimal $\mathbf{v}_{kis}$ of (40) satisfying

$$\mathbf{D}_{is} \bar{\mathbf{U}}_{is} \mathbf{v}_{kis} = \mathbf{u}_{kiis} \quad (41)$$

exists if and only if $\mathbf{D}_{is}$ (i.e., $\mathbf{Q}_{is}^T$) has a maximum row rank of $L$ (i.e., $N_c \leq L$). Consequently, the objective function of (40) will be $N|\mathbf{q}_{kiis}^T \mathbf{v}_{kis}| = N|\mathbf{u}_{kiis}^H \mathbf{u}_{kiis}| = N$ which scales with $N$ as desired. From these discussions, one can also understand that $\mathbf{x}_{mj}$ should also be selected in such a way that $\mathbf{Q}_{is}^T, \forall s, i$ are full row rank matrices.

## APPENDIX C
### PROOF OF *Theorem 2*

This appendix provides the SINR achieved by employing the existing and proposed approaches when $\mathbf{P}_{kji} = g_{kji}\mathbf{I}$ and $\mathbf{R}_{kji} = \mathbf{I}$. When cell $i$ has perfect CSI of its UEs, the SINR achieved by the $k$th UE in cell $i$ is given by

$$\gamma_{ki}^{pe} = \frac{\mathrm{E}\{|\mathbf{h}_{kiis}^H \mathbf{h}_{kiis}|^2\}}{\sum_{(m,j)\neq(k,i)}^{N_c} \mathrm{E}\{|\mathbf{h}_{kiis}^H \mathbf{h}_{mjis}|^2\} + \mathrm{E}\{|\mathbf{h}_{kiis}^H \tilde{\mathbf{w}}_{is}|^2\}}$$
$$= N \frac{L g_{kii}}{\sum_{m=1}^{\tilde{K}} \sum_{j=1,(m,j)\neq(k,i)}^{N_c} L g_{kji} + \sigma^2} \quad (44)$$

where we have used the fact that $\mathrm{E}\{\bar{\mathbf{h}}_{kiin} \bar{\mathbf{h}}_{kiin}^H\} = g_{kii}\mathbf{I}$. With the proposed approach, we will have the following SINR

$$\gamma_{kis}^{pr} = \frac{\mathrm{E}\{|\mathbf{h}_{kiis}^H \mathbf{a}_{kiis}|^2\}}{\sum_{(m,j)\neq(k,i)} \mathrm{E}\{|\mathbf{h}_{mjis}^H \mathbf{a}_{kiis}|^2\} + \mathrm{E}\{|\tilde{\mathbf{w}}_i^H \mathbf{a}_{kiis}|^2\}}. \quad (45)$$

Using (34) and (35), one can express $\mathrm{E}\{|\mathbf{h}_{mjis}^H \mathbf{a}_{kiis}|^2\}$ and $\mathrm{E}\{|\mathbf{w}_i^H \mathbf{a}_{kiis}|^2\}$ as

$$\mathrm{E}\{|\mathbf{h}_{mjis}^H \mathbf{a}_{kiis}|^2\} = \mathbf{v}_{kis}^H \Bigg( \mathbf{X}_{mj}^*[\mathbf{C}_{mji}\mathbf{f}_s^*\mathbf{f}_s^T \mathbf{C}_{mji}$$
$$+ \sum_{u=1}^{\tilde{K}} \sum_{v=1,(u,v)\neq(m,j)}^{N_c} \mathbf{C}_{mjuvis}]\mathbf{X}_{mj}^T + \sigma^2 \mathrm{tr}\{\mathbf{C}_{mjis}\}\mathbf{I} \Bigg) \mathbf{v}_{kis}$$
$$= N \mathbf{v}_{kis}^H \Bigg( \mathbf{X}_{mj}^*[N g_{mji}^2 \mathbf{f}_s^* \mathbf{f}_s^T + \sum_{u=1}^{\tilde{K}} \sum_{v=1,(u,v)\neq(m,j)}^{N_c}$$
$$L g_{mji} g_{uvi} \mathbf{I}] \mathbf{X}_{mj}^T + L\sigma^2 g_{mji} \mathbf{I} \Bigg) \mathbf{v}_{kis}$$

$$\mathrm{E}\{|\mathbf{w}_i^H \mathbf{a}_{kiis}|^2\} =$$
$$\sigma^2 \mathbf{v}_{kis}^H (\sum_{m=1}^{\tilde{K}} \sum_{j=1}^{N_c} \mathbf{X}_{mj}^* \mathbf{C}_{mji} \mathbf{X}_{mj}^T + \sigma^2 N \mathbf{I}) \mathbf{v}_{kis} \quad (46)$$
$$= N\sigma^2 \mathbf{v}_{kis}^H (\sum_{m=1}^{\tilde{K}} \sum_{j=1}^{N_c} g_{mji} \mathbf{X}_{mj}^* \mathbf{X}_{mj}^T + \sigma^2 \mathbf{I}) \mathbf{v}_{kis}.$$

By substituting (46) into (45), we will get $\gamma_{kis}^{pr}$ in (28). Furthermore, when $\gamma_{kis}^{pe} \gg 1$ and $\gamma_{kis}^{pr} \gg 1$, we will have

$$R_{kis}^{pe} - R_{kis}^{pr} = \log_2(1 + \gamma_{kis}^{pe}) - \log_2(1 + \gamma_{kis}^{pr})$$
$$= \log_2 \left( \frac{\gamma_{kis}^{pe}}{\gamma_{kis}^{pr}} \right).$$

It follows that $\frac{\gamma_{kis}^{pe}}{\gamma_{kis}^{pr}}$ is upper bounded by (42). By employing this inequality, we get the bound and approximation in (31).

## APPENDIX D
### PROOF OF *Lemma 1*

This appendix provides the proof of *Lemma 1*. From (9), we obtain

$$\hat{d}_{kis} = \mathbf{a}_{kiis}^H (\mathbf{h}_{kiis} d_{kis} + \sum_{(m,j)\neq(k,i)} \mathbf{h}_{mjis} d_{mjs} + \tilde{\mathbf{w}}_{is}).$$

Now if we employ the existing MMSE channel estimation and under high SNR regime, we will have $\mathbf{a}_{kiis} \approx \mathbf{h}_{kiis} + \sum_{j=1,j\neq i}^{N_c} \mathbf{h}_{kjis}$. By substituting this $\mathbf{a}_{kiis}$ into the above equation and after some manipulations, the SINR of $d_{kis}$ ($\gamma_{kis}^{ex}$) can be expressed as

$$\gamma_{kis}^{ex} = \frac{\mathrm{E}\{|\sum_{j=1}^{N_c} \mathbf{a}_{kiis}^H \mathbf{h}_{kiis}|^2\}}{\sum_{(m,j)\neq(k,i)}^{N_c} \mathrm{E}\{|\sum_{v=1}^{N_c} \mathbf{a}_{kii}^H \mathbf{h}_{mjis}|^2\} + \mathrm{E}\{|\mathbf{a}_{kiis}^H \tilde{\mathbf{w}}_{is}|^2\}}$$
$$= \frac{N^2 L^2 g_{kii}^2 + NL \sum_{j=1,j\neq i}^{N_c} g_{kii} g_{kji}}{\sum_{j=1,j\neq i}^{N_c} N^2 L^2 g_{kji}^2 + NL\sigma^2 \sum_{j=1}^{N_c} g_{kji} + NL\kappa}$$
$$= \frac{NL g_{kii}^2 + \sum_{j=1,j\neq i}^{N_c} g_{kii} g_{kji}}{\sum_{j=1,j\neq i}^{N_c} NL g_{kji}^2 + \sigma^2 \sum_{j=1}^{N_c} g_{kji} + \kappa}$$

where $\kappa = \sum_{m=1}^{\tilde{K}} \sum_{j=1,m\neq k,(m,j)\neq(k,i)}^{N_c} \sum_{u=1}^{N_c} g_{mji} g_{kui}$. The second equality is due to the fact that $\mathrm{E}\{|\mathbf{a}_{kiis}^H \tilde{\mathbf{w}}_{is}|^2\} = NL\sigma^2 \sum_{j=1}^{N_c} g_{kji}$. When $\sigma^2 \to 0$, we will get

$$\gamma_{kis}^{ex} \approx \frac{NL g_{kii}^2 + \sum_{j=1,j\neq i}^{N_c} g_{kii} g_{kji}}{\sum_{j=1,j\neq i}^{N_c} NL g_{kji}^2 + \kappa} \geq \frac{NL}{NLN_c + \tilde{K}N_c^2} \quad (47)$$

where the second inequality follows from the fact that $g_{kii} \geq g_{mji}, j \neq i$.

Using (28), the lower bound $\gamma_{kis}^{pr}$ becomes (43). Furthermore, scaling $\mathbf{v}_{kis}$ obtained from (11) with an arbitrary scalar value does not affect $\gamma_{kis}^{pr}$. Thus, one can rescale $\mathbf{v}_{kis}$ in a way that $\max_{m,j} L\mathbf{v}_{kis}^H \mathbf{X}_{mj}^* \mathbf{X}_{mj}^T \mathbf{v}_{kis} = 1, (m,j) \neq (k,i)$. With this scaling and $\sigma^2 \to 0$, we will have

$$\gamma_{kis}^{pr} \approx \frac{N g_{kii}^2}{\kappa} \geq \frac{N}{\tilde{K}^2 N_c^2} \quad (48)$$

where the second in equality follows from the fact that $g_{kii} \geq g_{mji}, j \neq i$. By employing (47) and (48), one can obtain $\gamma_{kis}^{pr} \geq \gamma_{kis}^{ex}$ when

$$N \geq \frac{N_c \tilde{K}(\tilde{K}L - 1)}{L}. \quad (49)$$

## APPENDIX E
### RELATION BETWEEN $L$, $T_d$ AND $B$

This appendix provides a brief description on the relations between $L$, $T_d$ and $B$. Since these rations do not depend on the number of cells, antennas and UEs, we consider a system having one transmitter and receiver both equipped with a single antenna. If we transmit an arbitrary waveform $x(t)$

$$\frac{\gamma_{kis}^{pe}}{\gamma_{kis}^{pr}} = \frac{Lg_{kii}\mathbf{v}_{kis}^H\left(\sum_{(m,j)\neq(k,i)}\left[\sum_{(u,v)\neq(m,j)}Lg_{mji}g_{uvi}\mathbf{X}_{mj}^*\mathbf{X}_{mj}^T + L\sigma^2 g_{mji}\mathbf{I}\right] + \sigma^2\tilde{\mathbf{Z}}\right)\mathbf{v}_{kis}}{(\sum_{m=1}^{\tilde{K}}\sum_{j=1,(m,j)\neq(k,i)}^{N_c}Lg_{kji}+\sigma^2)(g_{kii}^2 + \frac{1}{N}\mathbf{v}_{kis}^H\left(\sum_{(u,v)\neq(k,i)}Lg_{kii}g_{uvi}\mathbf{X}_{ki}^*\mathbf{X}_{ki}^T + L\sigma^2 g_{kii}\mathbf{I}\right)\mathbf{v}_{kis})}$$

$$\leq \frac{L\mathbf{v}_{kis}^H\left(\sum_{(m,j)\neq(k,i)}\left[\sum_{(u,v)\neq(m,j)}Lg_{mji}g_{uvi}\mathbf{X}_{mj}^*\mathbf{X}_{mj}^T + L\sigma^2 g_{mji}\mathbf{I}\right] + \sigma^2\tilde{\mathbf{Z}}\right)\mathbf{v}_{kis}}{(\sum_{m=1}^{\tilde{K}}\sum_{j=1,(m,j)\neq(k,i)}^{N_c}Lg_{kji}+\sigma^2)g_{kii}} = \eta \qquad (42)$$

$$\gamma_{kis}^{pr} = \frac{Ng_{kii}^2 + \mathbf{v}_{kis}^H\left(\sum_{(u,v)\neq(k,i)}Lg_{kii}g_{uvi}\mathbf{X}_{ki}^*\mathbf{X}_{ki}^T + L\sigma^2 g_{kii}\mathbf{I}\right)\mathbf{v}_{kis}}{\mathbf{v}_{kis}^H\left(\sum_{(m,j)\neq(k,i)}\left[\sum_{(u,v)\neq(m,j)}Lg_{mji}g_{uvi}\mathbf{X}_{mj}^*\mathbf{X}_{mj}^T + L\sigma^2 g_{mji}\mathbf{I}\right] + \sigma^2\tilde{\mathbf{Z}}\right)\mathbf{v}_{kis}}\bigg|_{\mathbf{v}_{kis}=\text{soln. of (11)}}$$

$$\geq \frac{Ng_{kii}^2}{\mathbf{v}_{kis}^H\left(\sum_{(m,j)\neq(k,i)}\left[\sum_{(u,v)\neq(m,j)}Lg_{mji}g_{uvi}\mathbf{X}_{mj}^*\mathbf{X}_{mj}^T + L\sigma^2 g_{mji}\mathbf{I}\right] + \sigma^2\tilde{\mathbf{Z}}\right)\mathbf{v}_{kis}}\bigg|_{\mathbf{v}_{kis}=\text{soln. of (11)}} \qquad (43)$$

having bandwidth $B$ with carrier frequency $f_c$, the received signal can be expressed as

$$y(t) = \sum_i a_i(t)x(t - \tau_i(t)) + w(t) \qquad (50)$$

where $a_i(t)$ and $\tau_i(t)$ are the overall attenuation and propagation delay at time t from the transmitter to the receiver in path $i$, respectively. And its discrete time equivalent model becomes

$$y[m] = \sum_l h_l[m]x[m-l] + w[m] \qquad (51)$$

where $h_l[m]$ is the $l$th (complex) channel filter tap at the $m$th symbol period $T_s = \frac{1}{B}$, $w[m]$ is the low-pass filtered noise at the sampling instant $mT_s$, $x[m]$ is the $m$th transmitted sample at time $mT_s$.

When $\tau_i$ and $a_i$ are time invariant (which is the assumption of the paper), we can express $h_l$ as

$$h_l = \sum_i a_i^b g[l - \tau_i B] \qquad (52)$$

where $a_i^b = a_i e^{-j2\pi f_c \tau_i}$ and $g[.]$ is the pulse shaping filter which could be raised cosine, sinc or rect function [20], with $a_i$ and $\tau_i$ are the overall attenuation and propagation delay from the transmitter and receiver on path $i$, respectively.

The coherence time of the channel $T_c$ is the interval over which $h_l[m]$ changes significantly with m which is roughly approximated as [20]

$$T_c = \frac{\alpha}{D_s}, \quad D_s = \max_{i,j} f_c|\tau_i - \tau_j| \qquad (53)$$

where $1 \leq \alpha \leq 8$.

The multipath delay spread $T_d$ is defined as the difference in the propagation time between the strongest and shortest path having a significant energy, i.e.,

$$T_d = \max_{i,j}|\tau_i - \tau_j|. \qquad (54)$$

And the coherence bandwidth, which explains how fast the channel changes in frequency, is therefore given by

$$W_c = \frac{1}{T_d}. \qquad (55)$$

The channel coherence bandwidth $W_c$ can also be treated as the bandwidth in which the channel is treated as 'almost constant'. Now when the bandwidth of the transmitted signal $B$ is considerably less than $W_c$, the wireless channel is termed as a flat fading channel (i.e., only one tap channel is required). However, when $B \gg W_c$, the wirelss channel is termed as a frequency selective channel (or wideband channel) where the number of taps is roughly given as $L = \frac{B}{W_c} = \frac{T_d}{T_s}$ [16]. In fact, this is reasonable as the channel condition changes every $W_c$ and at least $L$ channel taps are required to learn the full CSI information of the $B$ Hz bandwidth channel. From the discussions of this appendix, one can notice the following points

1) The channel coherence time $T_c$ depends on the carrier frequency $f_c$ and the channel propagation delay $\tau_i$
2) The channel propagation delay $\tau_i$ depends on different factors such as the number of local scatterers, the relative motions of the transmitter, receiver and scatters etc. For this reason, $T_d$ is fairly constant for a given wireless environment.
3) For a given environment, $L$ can be increased just by increasing $B$ (i.e., decreasing $T_s$).